\documentclass[prd,notitlepage,nofootinbib,superscriptaddress]{revtex4-1}

\usepackage{amsmath}
\usepackage{amsfonts}
\usepackage[utf8]{inputenc}
\usepackage{graphicx}
\usepackage{hyperref}
\hypersetup{colorlinks, linkcolor = [rgb]{0,0.0,0.75}, citecolor = [rgb]{0,0.0,0.75}, urlcolor = [rgb]{0,0.0,0.75}}

\makeatletter
\g@addto@macro\bfseries{\boldmath}
\makeatother
\begin{document}

\title{Excited-state effects in nucleon structure on the lattice using hybrid interpolators}

\author{Jeremy~R.~Green}
\email{jeremy.green@desy.de}
  \affiliation{NIC, Deutsches Elektronen-Synchroton, 15738 Zeuthen, Germany}

\author{Michael~Engelhardt}
  \affiliation{Department of Physics, New Mexico State University, Las Cruces, NM 88003-8001, USA}

\author{Nesreen~Hasan}
  \affiliation{Bergische Universität Wuppertal, 42119 Wuppertal, Germany}
  \affiliation{IAS, Jülich Supercomputing Centre, Forschungszentrum Jülich, 52425 Jülich, Germany}

\author{Stefan~Krieg}
  \affiliation{Bergische Universität Wuppertal, 42119 Wuppertal, Germany}
  \affiliation{IAS, Jülich Supercomputing Centre, Forschungszentrum Jülich, 52425 Jülich, Germany}

\author{Stefan~Meinel}
  \affiliation{Department of Physics, University of Arizona, Tucson, AZ 85721, USA}
  \affiliation{RIKEN BNL Research Center, Brookhaven National Laboratory, Upton, NY 11973, USA}

\author{John~W.~Negele}
  \affiliation{Center for Theoretical Physics, Massachusetts Institute of Technology, Cambridge, MA 02139, USA}

\author{Andrew~V.~Pochinsky}
  \affiliation{Center for Theoretical Physics, Massachusetts Institute of Technology, Cambridge, MA 02139, USA}

\author{Sergey~N.~Syritsyn} 
  \affiliation{RIKEN BNL Research Center, Brookhaven National Laboratory, Upton, NY 11973, USA}
   \affiliation{Department of Physics and Astronomy, Stony Brook University, Stony Brook, NY 11794, USA}

\date{\today}
\begin{abstract}
  It would be very useful to find a way of reducing excited-state
  effects in lattice QCD calculations of nucleon structure that has a
  low computational cost. We explore the use of hybrid interpolators,
  which contain a nontrivial gluonic excitation, in a variational
  basis together with the standard interpolator with tuned smearing
  width. Using the clover discretization of the field strength tensor,
  a calculation using a fixed linear combination of standard and
  hybrid interpolators can be done using the same number of quark
  propagators as a standard calculation, making this a cost-effective
  option. We find that such an interpolator, optimized by solving a
  generalized eigenvalue problem, reduces excited-state contributions
  in two-point correlators. However, the effect in three-point
  correlators, which are needed for computing nucleon matrix elements,
  is mixed: for some matrix elements such as the tensor charge,
  excited-state effects are suppressed, whereas for others such as the
  axial charge, they are enhanced. The results illustrate that the
  variational method is not guaranteed to reduce the net contribution
  from excited states except in its asymptotic regime, and suggest
  that it may be important to use a large basis of interpolators
  capable of isolating all of the relevant low-lying states.
\end{abstract}

\maketitle

\section{Introduction}

One of the most challenging sources of systematic uncertainty faced by
lattice QCD calculations of nucleon structure is excited-state
contamination: the failure to isolate the ground-state nucleon from
the tower of higher-energy states to which the interpolating operator
can couple. Although the unwanted excited states can be exponentially
suppressed by Euclidean time evolution, this is hindered by an
exponentially decaying signal-to-noise ratio~\cite{Lepage:1989hd} that
makes it impractical to evolve long enough in Euclidean time.

The variational method~\cite{Michael:1985ne, Luscher:1990ck,
  Blossier:2009kd} provides a way of improving the interpolating
operator such that the lowest-lying excited states can be
systematically removed. Variational approaches have been used to study
nucleon structure in Refs.~\cite{Engel:2009nh, Owen:2012ts,
  Yoon:2016dij, Dragos:2016rtx, Stokes:2018emx}, which used bases of
interpolators with different smearing widths and different site-local
spin structures, and Ref.~\cite{Egerer:2018xgu}, which used the
distillation method to enable the use of interpolators with a variety
of local structures including covariant derivatives. In these cases,
the variational setup was more computationally expensive than a
standard calculation because of the need for additional quark
propagators with different smeared sources\footnote{The comparison is
  more difficult when using the distillation method, which uses
  timeslice sources rather than the point sources used in standard
  calculations.}.

In this paper, we present a study of a variational setup that requires
the same number of quark propagators as a standard calculation, for a
fixed choice of optimized interpolator. This is accomplished by
supplementing the standard interpolator with hybrid
ones~\cite{Dudek:2012ag} that contain a gluonic
excitation\footnote{The resulting basis of interpolators is similar to
  the one called $\mathcal{B}_3$ in Ref.~\cite{Egerer:2018xgu}.}. In
Ref.~\cite{Dudek:2012ag}, it was found that the latter have the
next-largest overlaps onto the ground state, after the standard
interpolator. The use of hybrid interpolators presents the possibility
of an improvement over the standard approach at low computational
cost.

It should be stressed that this study, along with all previous ones,
uses only local interpolating operators, which are poor at isolating
multiparticle states. In practice, the true spectrum that includes
$N\pi$ and $N\pi\pi$-like states is not identified, meaning that the
calculation is not in a regime where the variational method has been
proven to improve the isolation of the ground state. Therefore, the
question of whether one interpolator is an improvement over another is
an empirical one, to be decided by examining excited-state
contributions in estimators for a variety of observables.

This paper is organized as follows. Section~\ref{sec:setup} discusses
our lattice setup, the basis of interpolating operators, and two
tuning runs. Results for two-point correlators are discussed in
Section~\ref{sec:two_point}, forward matrix elements are shown in
Section~\ref{sec:forward_me} and form factors are presented in
Section~\ref{sec:form_factors}. Our conclusions are given in
Section~\ref{sec:conclusions}.

\section{Lattice setup}
\label{sec:setup}

As this is an exploratory calculation, we use a single lattice
ensemble with a coarse lattice spacing at a heavier-than-physical pion
mass and a relatively small box size; its parameters are summarized in
Table~\ref{tab:lat_setup}. This has $2+1$ flavors of tree-level
improved Wilson-clover quarks coupled to the gauge links via two
levels of HEX smearing~\cite{Durr:2010aw}.

\begin{table*}
\centering
\begin{tabular}{c c c c | c c c c | c c c c}
Size & $\beta$ & $am_{ud}$ & $am_s$
& $a$ [fm] & $am_\pi$ & $m_\pi$ [MeV] & $m_\pi L$
& $T/a$ & $N_\mathrm{conf}$ & $N_\text{src}$ & $N_\mathrm{meas}$\\
\hline
  $24^3\times 48$ & 3.31 & $-0.09530$ & $-0.04$
  & 0.1163(4) & 0.1499(7) & 254(1) & 3.6
  & \{6, 8, 10\} & 600 & 48 & 57600
\end{tabular}
\caption{Parameters of the ensemble and measurements used in this
  work. The lattice spacing is taken from Ref.~\cite{Durr:2010aw},
  where it is set using the mass of the $\Omega$ baryon at the
  physical point. $N_\mathrm{conf}$ refers to the number of gauge
  configurations analyzed and
  $N_\mathrm{meas} = 2\times N_\mathrm{conf} \times N_\mathrm{src}$ is
  the number of measurements performed. The factor of 2 in
  $N_\mathrm{meas}$ accounts for the use of forward- and
  backward-propagating states.}
\label{tab:lat_setup}
\end{table*}

Aside from the interpolating operator, the methodology used here for
computing nucleon matrix elements and form factors is unchanged from
previous work such as Ref.~\cite{Green:2014xba}; the reader is
referred to that earlier work for details. Our focus is on seeing
whether excited-state contamination can be reduced, and therefore we
use three relatively short source-sink separations, $T$, ranging from
0.70 to 1.16~fm. We use two methods for determining matrix elements:
the ratio method --- for which the asymptotically leading
excited-state contributions decay as $e^{-\Delta E T/2}$, where
$\Delta E$ is the energy gap to the lowest excited state --- and the
summation method, for which they decay as $Te^{-\Delta E T}$.

Given a set of $N$ interpolating operators $\{\chi_i\}$, one would
like to find a linear combination $\chi=v_i\chi_i$ that has a reduced
coupling to excited states. The standard approach is to compute a
matrix of two-point correlators,
\begin{equation}
  C_{ij}(t) = \left\langle \chi_i^{\vphantom{\dagger}}(t) \chi_j^\dagger(0) \right\rangle,
\end{equation}
and then solve a generalized eigenvalue problem (GEVP)
\begin{equation}\label{eq:GEVP}
  C(t_2)v = \lambda C(t_1)v
\end{equation}
for some choice of $(t_1,t_2)$. It has been
shown~\cite{Blossier:2009kd} that by suitably increasing $t_1$ and
$t_2$ to remove contributions from higher excited states in the
determination of $v$, one can define improved estimators for the
ground-state energy and matrix elements, for which the leading
excited-state effect depends on the energy gap to state $N+1$ rather
than the second (i.e.\ first excited) state. However, it is known that
for light pion masses and large volumes, the number of low-lying
excited states with the quantum numbers of the nucleon is large due to
the presence of multiparticle ($N\pi$, $N\pi\pi$, etc.)
states~\cite{Tiburzi:2009zp, Tiburzi:2015tta, Bar:2015zwa,
  Bar:2016uoj, Hansen:2016qoz, Bar:2017kxh, Bar:2018xyi,
  Green:2018vxw}. Removing the effects of these states would require
that the basis include at least one operator for each state. In
addition, it has been found in meson spectroscopy calculations that
nonlocal multiparticle interpolators must be included in order to
correctly identify the multiparticle spectrum: see, e.g.,
Ref.~\cite{Dudek:2012xn}. For nucleons, the need for nonlocal
operators is also supported by chiral perturbation theory, which
predicts at leading order that the ratio of couplings for single
nucleon and nucleon-pion states is the same for all local
operators~\cite{Bar:2015zwa}. This defeats the diagonalization
procedure of Eq.~\eqref{eq:GEVP} such that $N\pi$ states cannot be
removed.

The challenge of systematically removing all contributions from the
lowest-lying excited states will be left to future work. Instead, we
hope to find an improved local operator that can be used in existing
software with minimal modifications and with little additional
computational cost. Our standard operator is
\begin{equation}\label{eq:op_standard}
  \chi_1 = \epsilon_{abc}(\tilde u_a^T C\gamma_5P_+ \tilde d_b) \tilde u_c,
\end{equation}
where $P_+=(1+\gamma_4)/2$ is a positive parity (nonrelativistic)
projector and $\tilde q$ is a smeared quark field. When used in a
two-point or three-point correlator with a polarization matrix that
includes a factor of $P_+$, the projector is applied to all three
quarks. This allows for computational cost savings in the quark
propagators used for constructing correlators: only half of the
propagator solves are required~\cite{Gockeler:1995wg}. Of the three
possible site-local nucleon operators, $\chi_1$ is the only one that
can be constructed using positive-parity-projected quark fields (see
e.g.\ Appendices B and C of Ref.~\cite{Basak:2005ir}).

We also consider hybrid operators, introduced in
Ref.~\cite{Dudek:2012ag}, which include an insertion of the
chromomagnetic field $B_i=-\frac{1}{2}\epsilon_{ijk}F_{jk}$ and are
interpreted as having a nontrivial gluonic excitation. Using the
clover discretization of $F_{\mu\nu}$~\cite{Sheikholeslami:1985ij}, no
additional quark propagators are needed for constructing two-point
correlators using hybrid operators. Two different nucleon operators
exist that use positive-parity-projected quark fields:
\begin{align}
\chi_2 &= \epsilon_{abc}\left[ (B_i \tilde u)_a^T C\gamma_j P_+ \tilde d_b\right] \gamma_i\gamma_j \tilde u_c
 - \epsilon_{abc}\left[ \tilde u_a^T C\gamma_j P_+ (B_i \tilde d)_b\right] \gamma_i\gamma_j \tilde u_c,\\
\chi_3 &= \epsilon_{abc}\left[ (B_i \tilde u)_a^T C\gamma_j P_+ \tilde d_b\right] P_{ij} \tilde u_c
 - \epsilon_{abc}\left[ \tilde u_a^T C\gamma_j P_+ (B_i \tilde d)_b\right] P_{ij} \tilde u_c,
\end{align}
where $P_{ij}=\delta_{ij}-\frac{1}{3}\gamma_i\gamma_j$. These differ
in the spin of the three quarks: for $\chi_2$ it is $\frac{1}{2}$ and
for $\chi_3$ it is $\frac{3}{2}$. In both cases, this is combined with
the chromomagnetic field to produce an overall spin of $\frac{1}{2}$.

Our production strategy begins with two tuning runs where only
two-point correlators are computed. The first uses only $\chi_1$ and
serves to select the quark smearing parameters. The second serves for
determining the coefficients $v_i$ of an optimized operator
$\chi_\text{opt}=v_i^*\chi_i$. These are followed by a production run
with higher statistics in which both two-point and three-point
correlators are computed. The three-point correlators are computed
using the standard operator $\chi_1$ and the linear combination
$\chi_\text{opt}$, in both cases keeping the same operator at the
source and the sink. For three-point correlators, using
$\chi_\text{opt}$ at the source and the sink requires a different
sequential propagator than for $\chi_1$, but the total number of
propagators needed in each case is the same. We chose not to compute
all nine combinations of source and sink interpolators in three-point
correlators because this would require nine times as many sequential
propagators as a standard calculation.

\subsection{Tuning of quark smearing}

We use Wuppertal smearing~\cite{Gusken:1989qx},
$\tilde q\propto (1+\alpha H)^Nq$, where $H$ is the nearest-neighbor
gauge-covariant hopping matrix constructed using the same smeared
links used in the fermion action. The parameter $\alpha$ is fixed to
3.0, and $N$ is varied to produce different smearing widths. The
smearing radius is determined by taking a color field $\varphi(\vec x)$
with support only at the origin and then defining a density from the
squared norm of the smeared field:
$\rho(\vec x)=|\tilde\varphi(\vec x)|^2$. Finally, we take the
root-mean-squared radius:
\begin{equation}
  r^2 = \frac{\sum_{\vec x} |\vec x|^2\rho(\vec x)}{\sum_{\vec x}\rho(\vec x)}.
\end{equation}
For the tuning run, we used $N\in\{20,40,70,110,160\}$, which
correspond roughly to $r/a\in\{3,4,5,6,7\}$.

\begin{figure*}
  \includegraphics[width=\textwidth]{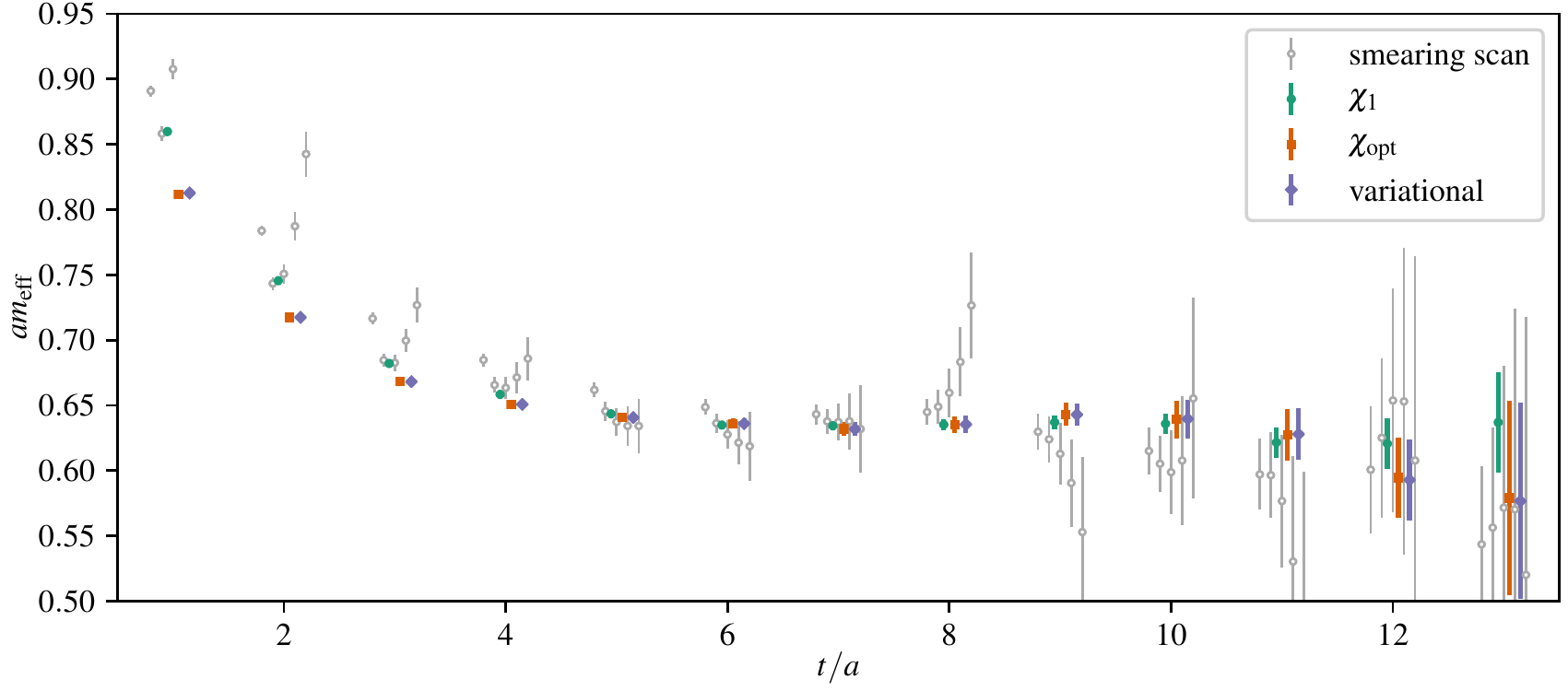}
  \caption{Effective mass of the nucleon. The five open gray circles
    at each $t$ show results from the low-statistics tuning of the
    quark smearing: the number of smearing steps increases from left
    to right. Filled symbols are from the full-statistics run: the
    standard operator (green circles), $\chi_\text{opt}$ chosen based
    on the second tuning run (orange squares), and the variationally
    optimized operator based on the full-statistics run (blue
    diamonds).}
  \label{fig:meff}
\end{figure*}

Figure~\ref{fig:meff} shows the effective mass
$am_\text{eff}(t)=\log\frac{C(t)}{C(t+a)}$ for each smearing
width. For $t=2a$ and $3a$, we can see that the minimum lies near
$N=40$ and $70$. Based on this, we decided to use the same smearing
parameter, $N=45$, that was used in a previous
calculation~\cite{Green:2012ej, Green:2012ud, Green:2014xba}.

\subsection{Tuning of variational operator}

The hybrid operators are constructed using a chromomagnetic field made
from smeared gauge links. We take the smeared links to which the
fermions couple in the action and then apply additional
three-dimensional stout smearing\footnote{We have not studied the
  effect of varying this.}~\cite{Morningstar:2003gk}: 20 steps with
$\rho=0.1$. The traceless part of the clover definition of the field
strength tensor is used.

\begin{figure}
  \includegraphics[width=0.5\textwidth]{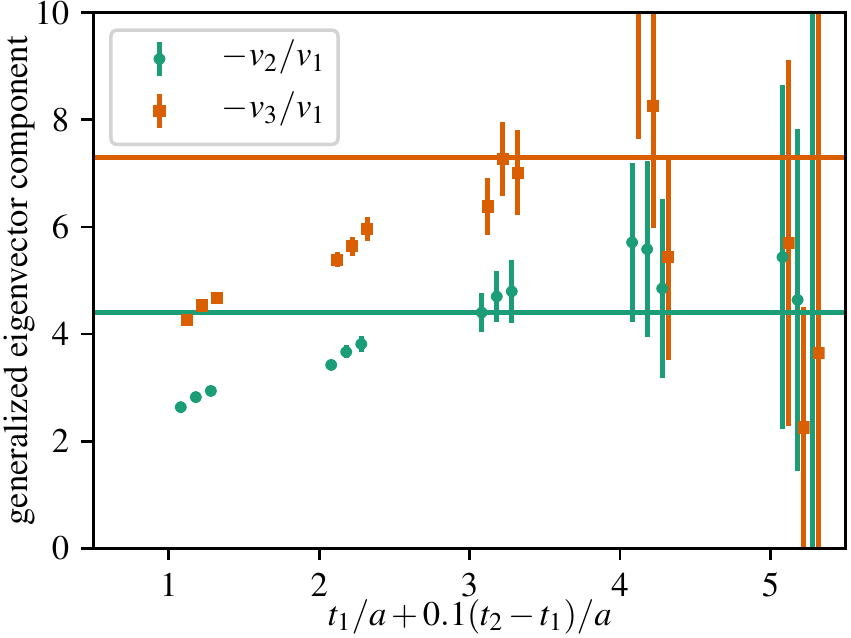}
  \caption{Components of the generalized eigenvector $v_i$, normalized
    to $v_1$. The GEVP was solved for $t_1/a\in[1,5]$ and
    $(t_2-t_1)/a\in[1,3]$. The horizontal lines indicate the values
    used in the optimized operator
    $\chi_\text{opt}=v_i\chi_i$, determined from the
    lower-statistics tuning run.}
  \label{fig:gevp_coeffs}
\end{figure}

For the second tuning run, we computed the full $3\times 3$ matrix of
two-point correlators. Solving the GEVP yielded the coefficients $v_i$
for $\chi_\text{opt}$; we did this at the largest available time
separations before the noise became too large. For our choice of
operators, the correlator matrix is real and thus $v_i$ are also real.
Normalizing such that $v_1=1$, we selected
$\chi_\text{opt}=\chi_1-4.4\chi_2-7.3\chi_3$. The determination of
coefficients from the subsequent full-statistics run is shown in
Fig.~\ref{fig:gevp_coeffs} for a range of $t_1$ and $t_2$ in the
GEVP. Our selection based on the tuning run is consistent with the
values determined at large times in the full run.

\section{Results}

\subsection{Two-point correlators}
\label{sec:two_point}

\begin{figure}
  \includegraphics[width=0.5\textwidth]{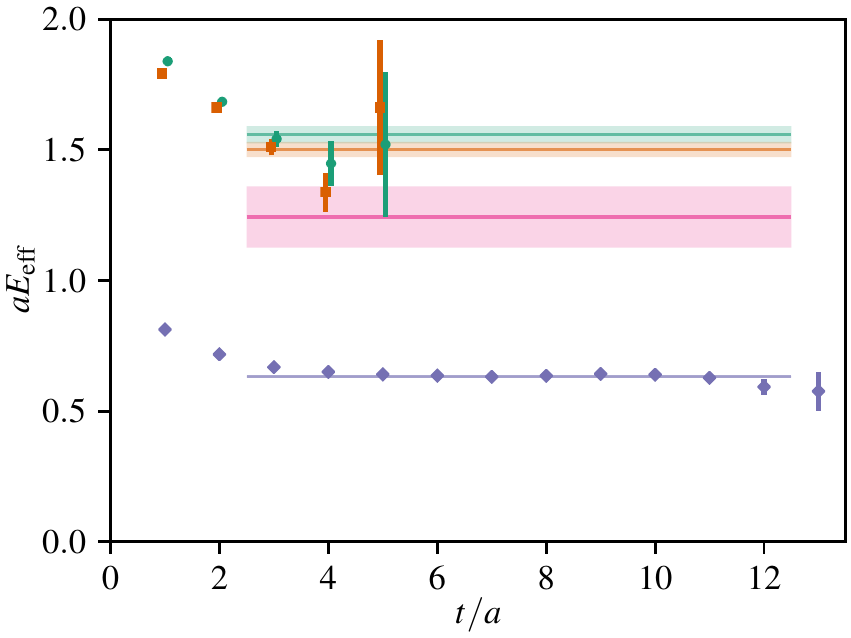}
  \caption{Effective energies for the three GEVP-projected correlators
    at zero momentum (in order of increasing energy: blue diamonds,
    orange squares, and green circles) and energies from the
    four-state fit (horizontal bands). The colors of the bands
    correspond to the effective energy with which they are identified
    in Section~\ref{sec:two_point}; the additional state has a magenta
    band. The horizontal range of the bands indicates the time
    separations in the two-point correlator matrix that are fitted.}
  \label{fig:variational}
\end{figure}

The full $3\times 3$ matrix of two-point correlators was computed in
the full-statistics run, allowing for a more detailed analysis.  We
begin by determining the excited energy levels using the variational
method. Solving the GEVP at $(t_1,t_2)=(3a,5a)$ yields an eigenvector
$v_n$ for each state $n$. This allows us to define projected
correlators $C_n(t)=v_n^\dagger C(t)v_n$ and then compute their
effective energies; these are shown in Fig.~\ref{fig:variational}. The
two excited energies are nearly degenerate and lie in the range from 1
to 1.5~GeV above the ground-state nucleon; this is similar to the
hybrid states observed in Ref.~\cite{Dudek:2012ag}. However, we stress
that this should not be considered a reliable determination of the
spectrum, as many lower-lying excitations are expected and can not be
identified using our small basis of three operators.

\begin{figure}
  \includegraphics[width=0.4\textwidth]{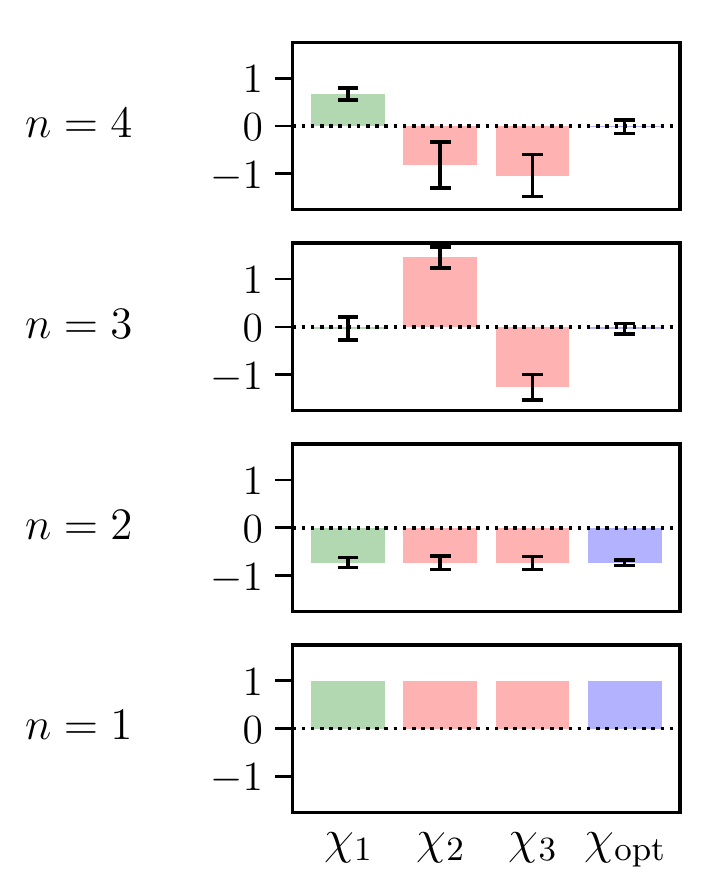}
  \caption{Overlap factors normalized to the ground state,
    $Z_{i,n}/Z_{i,1}$. These are shown for the three operators
    $\chi_i$ and for their linear combination $\chi_\text{opt}$.}
  \label{fig:overlaps}
\end{figure}

The presence of residual excited-state effects in the effective mass
(Fig.~\ref{fig:meff}) suggests that more than three states are needed
to describe the two-point correlators. We employ the fit model
\begin{equation}
  C_{ij}(t) = \sum_{n=1}^{N_\text{states}} Z_{i,n} Z_{j,n} e^{-E_n t},
\end{equation}
with the energies ordered $E_1<E_2<\cdots$. We obtained a good fit to
the range $t/a\in[3,12]$ using $N_\text{states}=4$, which yielded
$\chi^2/\text{dof}=0.89$ ($p=0.68$). The four energy levels are shown
in Fig.~\ref{fig:variational}. States 3 and 4 are consistent with the
effective energies from the GEVP-projected correlators for the two
excited states with $t/a=3$. The additional energy level, state 2,
sits below the two excited states identified by the GEVP.

This identification of states between the GEVP and the four-state fit
is also supported by Fig~\ref{fig:overlaps}, which shows the overlap
factors normalized to the ground state, $Z_{i,n}/Z_{i,1}$. It also
shows the normalized overlap factors for the operator
$\chi_\text{opt}$, which are determined via
$Z_{\text{opt},n}=v_i Z_{i,n}$. The fit results indicate that $\chi_1$
has a significant overlap with state 4, which is eliminated in
$\chi_\text{opt}$. The overlap of $\chi_1$ with state 3 is consistent
with zero, and this is preserved in $\chi_\text{opt}$ despite the
large overlaps of $\chi_2$ and $\chi_3$ with state 3. The operators
show no significant difference in the relative overlaps with states 1
and 2; because of this, the GEVP was insensitive to state 2 and was
unable to eliminate the coupling of $\chi_\text{opt}$ to it.

\begin{figure*}
  \includegraphics[width=\textwidth]{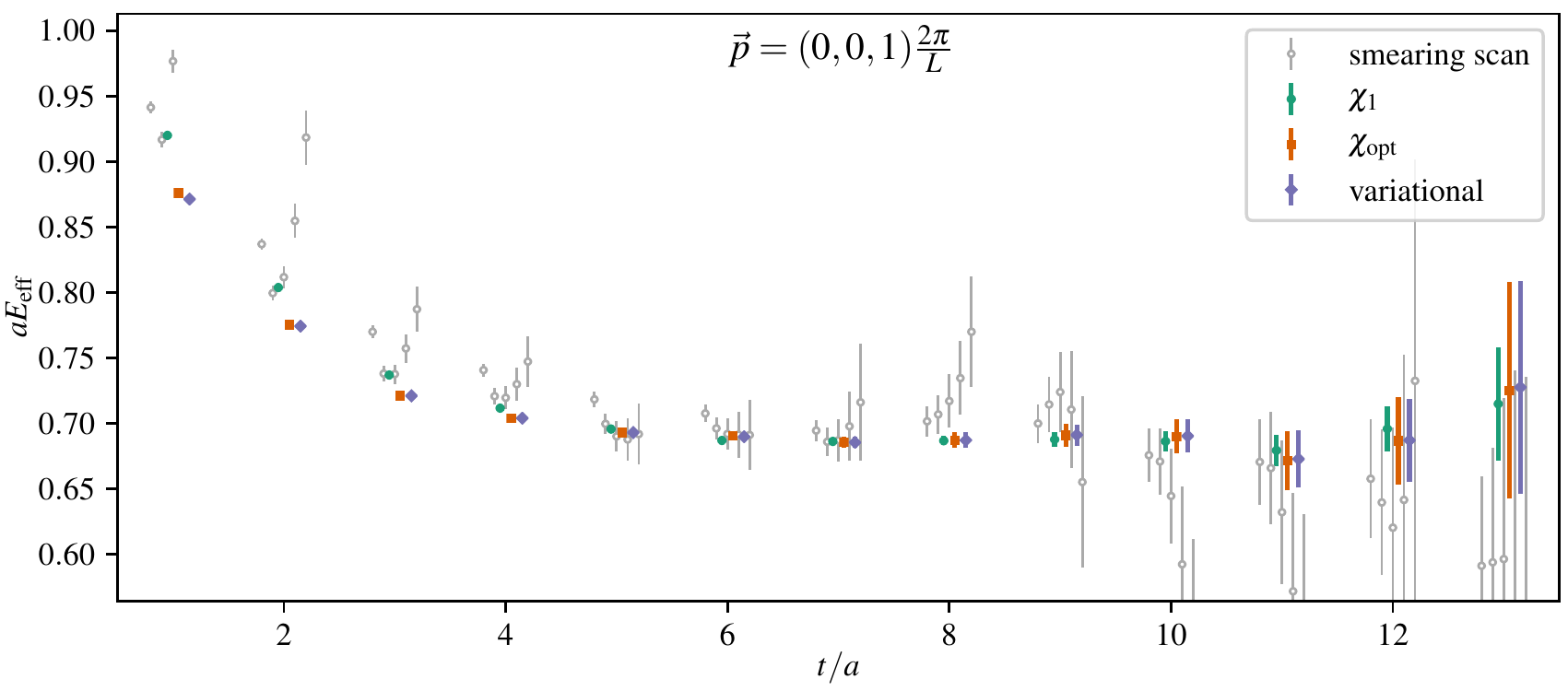}\\
  \includegraphics[width=\textwidth]{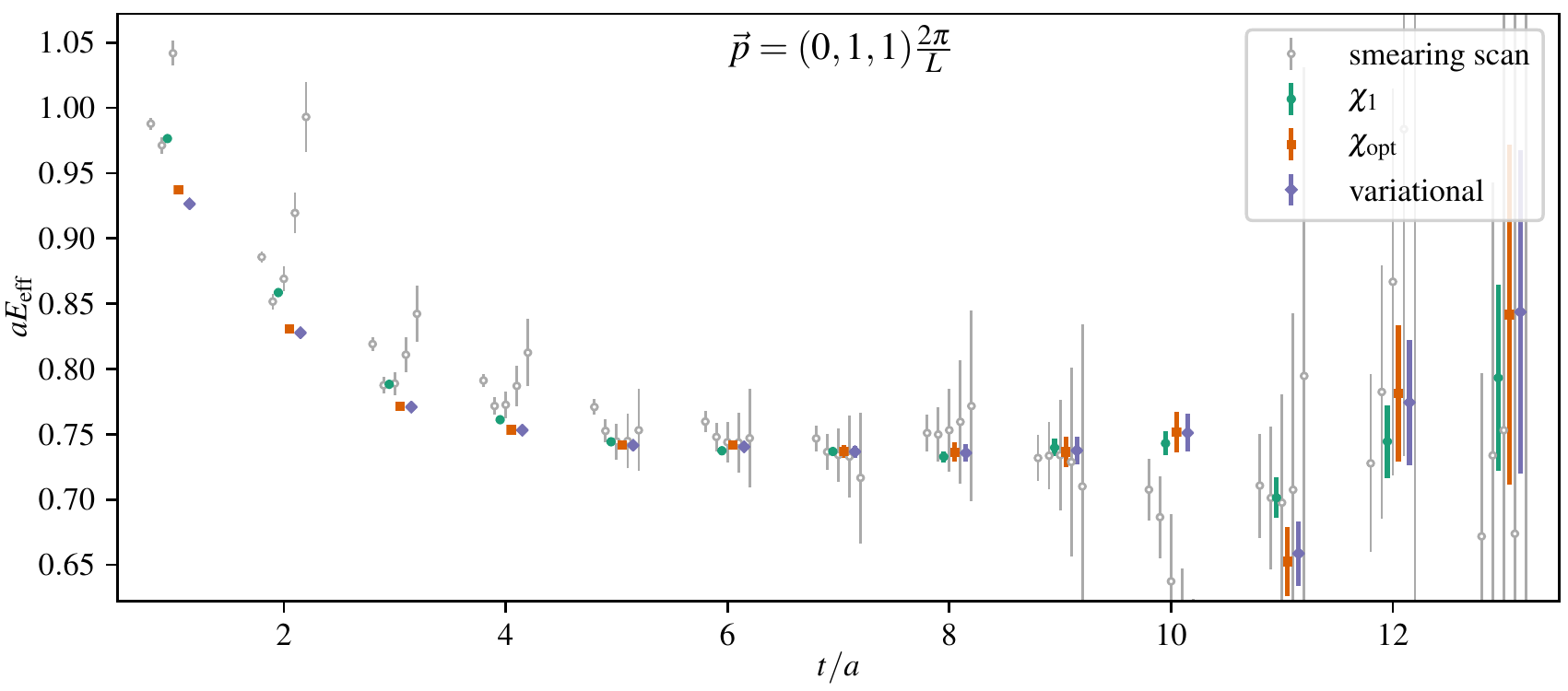}\\
  \includegraphics[width=\textwidth]{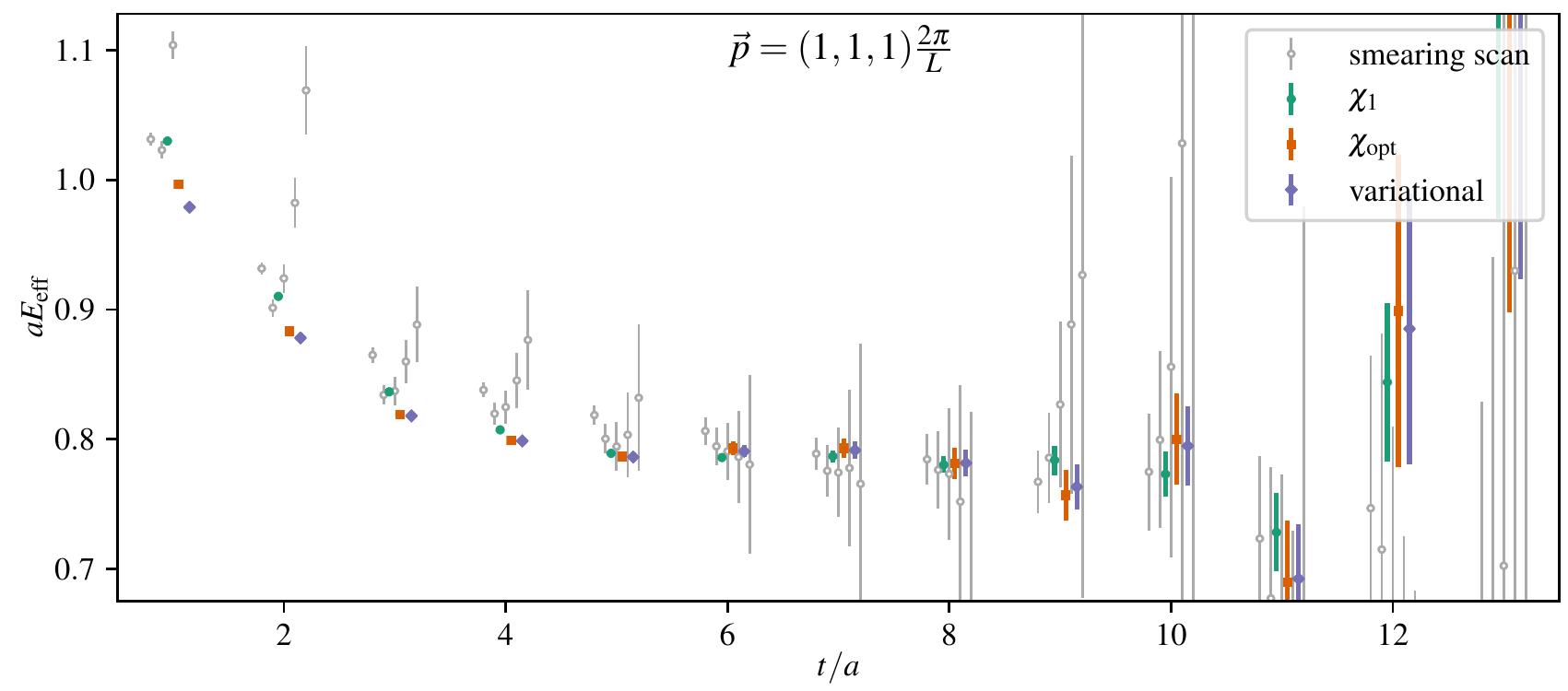}
  \caption{Effective energies of the nucleon at nonzero momentum. See
    the caption of Fig.~\ref{fig:meff}. Note that $\chi_1$ and
    $\chi_\text{opt}$ were tuned at zero momentum, whereas the full
    variational analysis is retuned at nonzero momentum.}
  \label{fig:Eeff}
\end{figure*}

Finally, we can compare effective energies produced using different
operators. In addition to $\chi_1$ and $\chi_\text{opt}$, which were
tuned at zero momentum using reduced statistics, we also perform a new
variational analysis based on the full-statistics run. In
Fig.~\ref{fig:meff}, there is no significant difference between
$\chi_\text{opt}$ and the new variational operator. Both of them have
smaller effective masses than $\chi_1$ at early times, indicating a
significant reduction in excited-state contributions. However, they
also suffer from increased statistical
uncertainty. Figure~\ref{fig:Eeff} shows the same comparison at
nonzero momentum. In all cases, $\chi_\text{opt}$ shows smaller
excited-state effects than $\chi_1$. At the smallest values of $t$,
the new variational operator also shows an improvement over
$\chi_\text{opt}$, and this effect grows with momentum. This is not
surprising, as the new variational operator is tuned for each
momentum, whereas $\chi_\text{opt}$ was chosen at $\vec p=0$.

\subsection{Forward matrix elements}
\label{sec:forward_me}

We have only computed two combinations of source and sink
interpolators in three-point correlators: those with the same
interpolator at the source and the sink, which is chosen to be either
$\chi_1$ or $\chi_\text{opt}$. For comparing these two setups, we
start by considering observables computed from isovector matrix
elements at zero momentum: the axial, tensor, and scalar charges
($g_A$, $g_T$, and $g_S$), and the average momentum fraction
$\langle x\rangle_{u-d}$.

\begin{figure*}
  \includegraphics[width=0.495\textwidth]{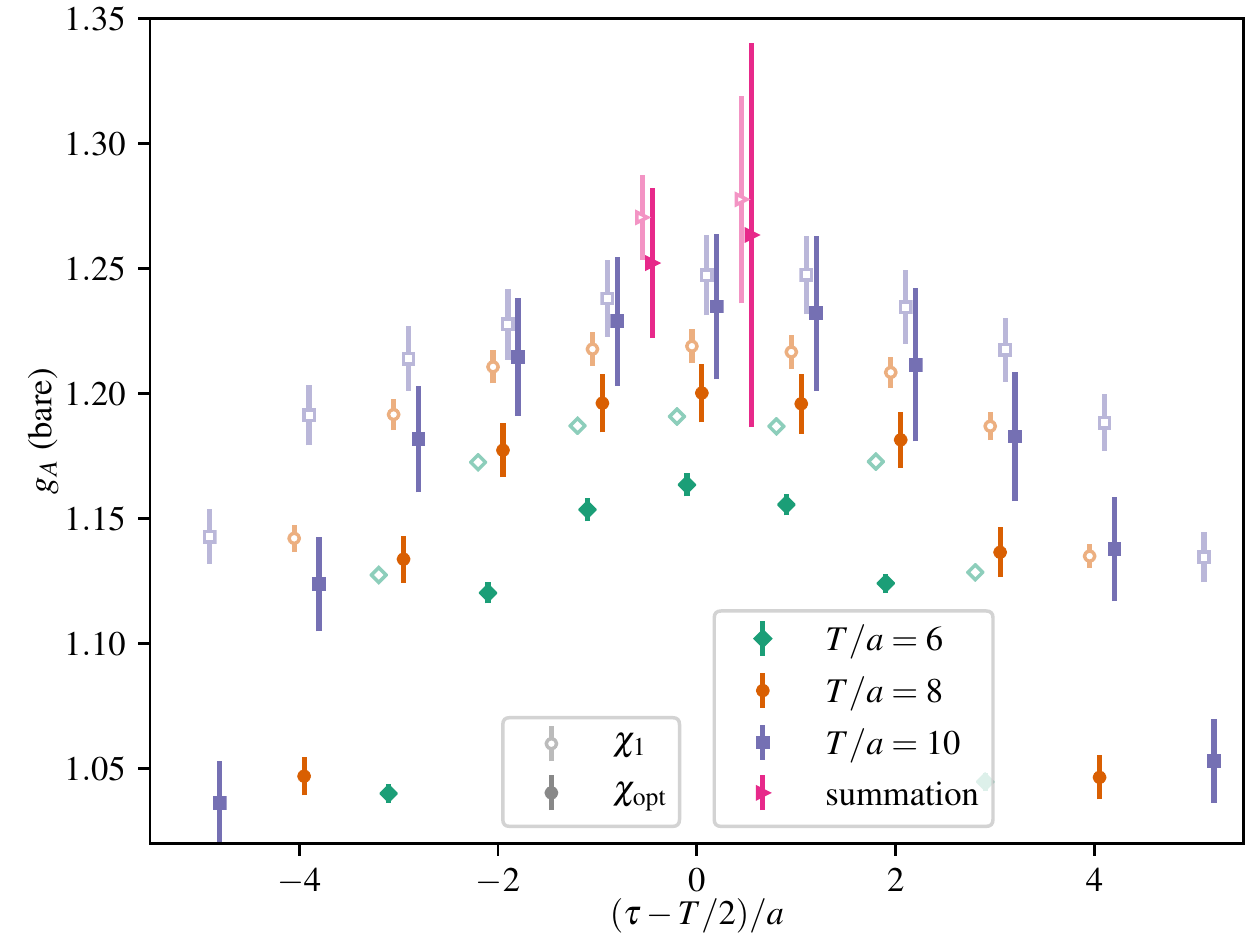}
  \includegraphics[width=0.495\textwidth]{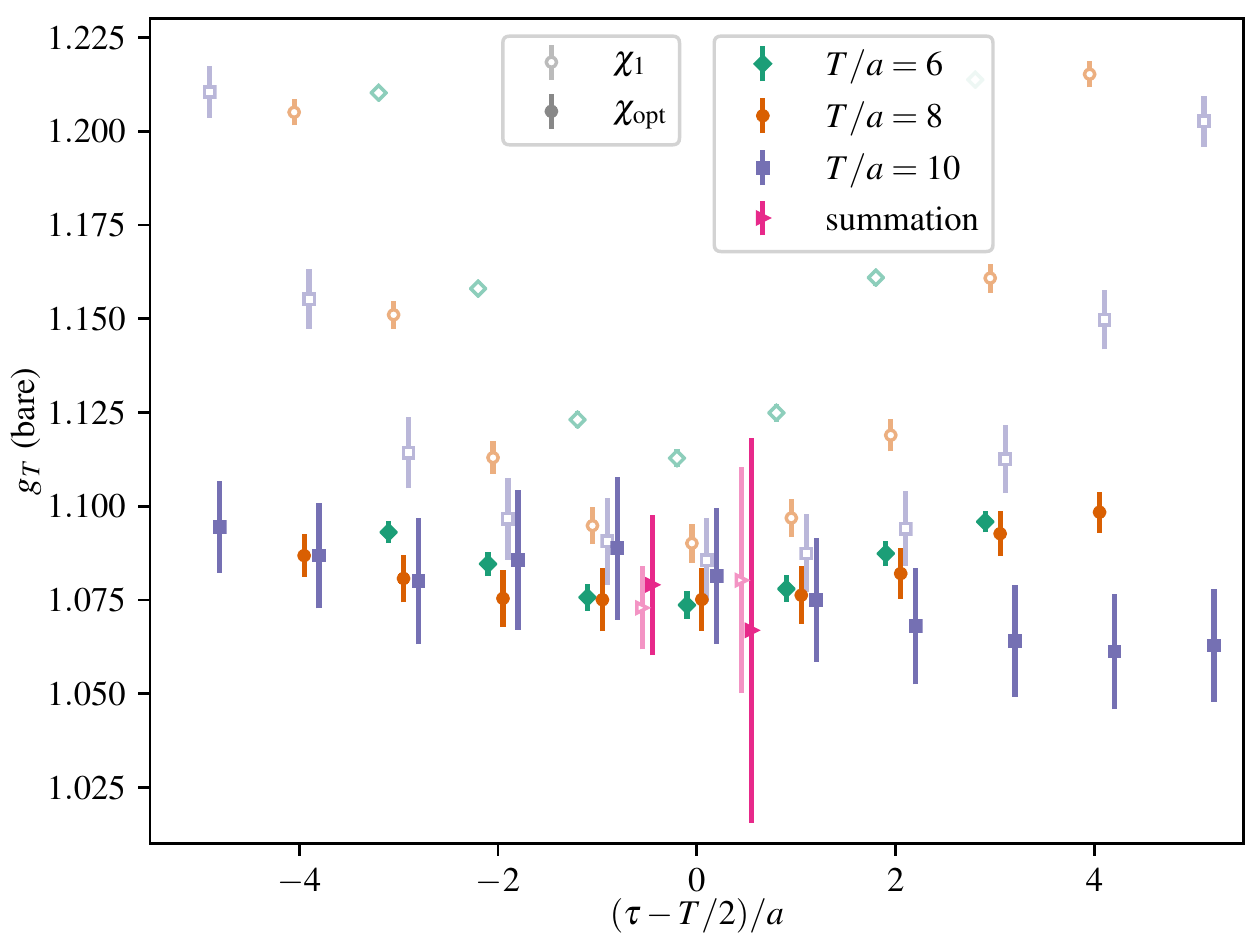}\\
  \includegraphics[width=0.495\textwidth]{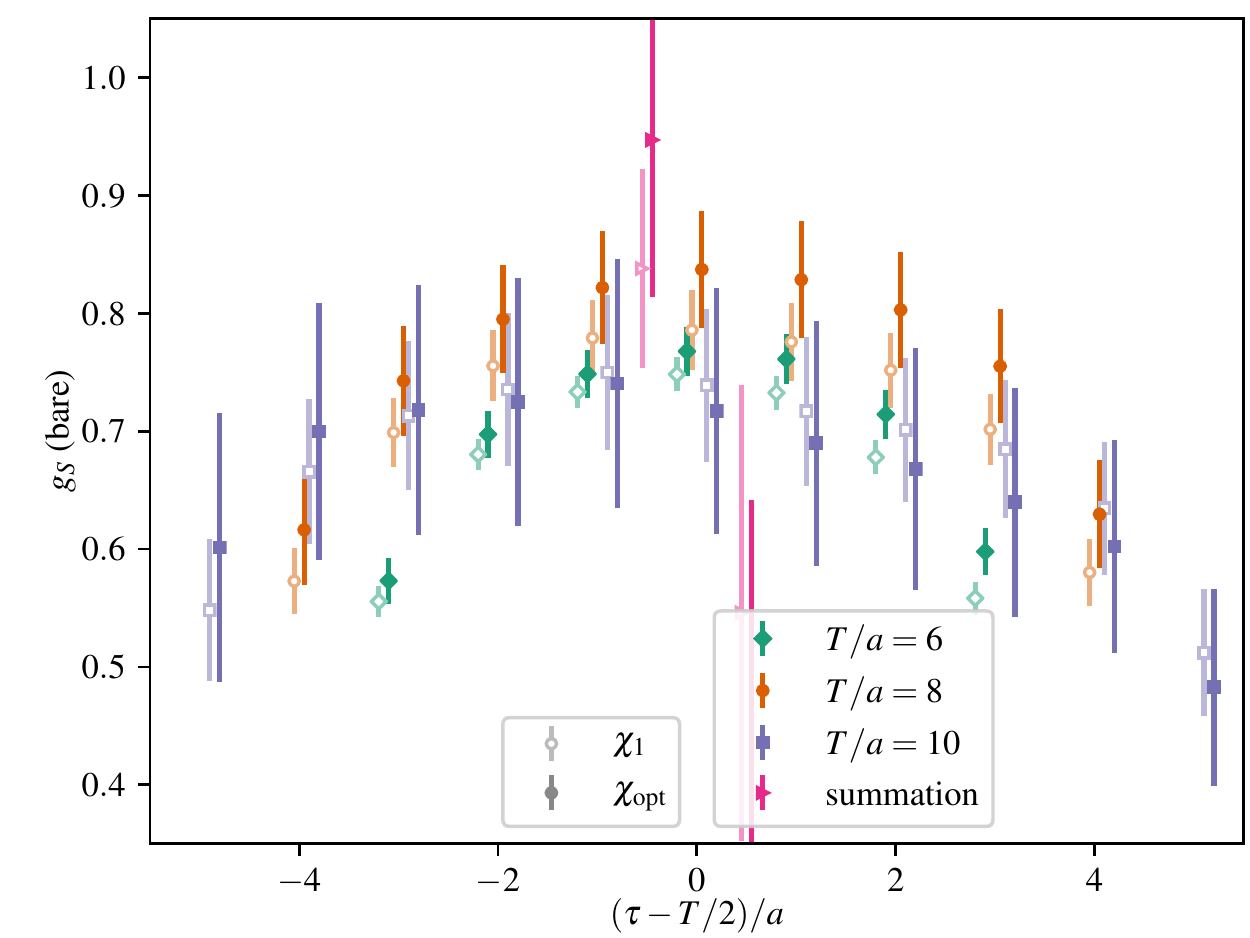}
  \includegraphics[width=0.495\textwidth]{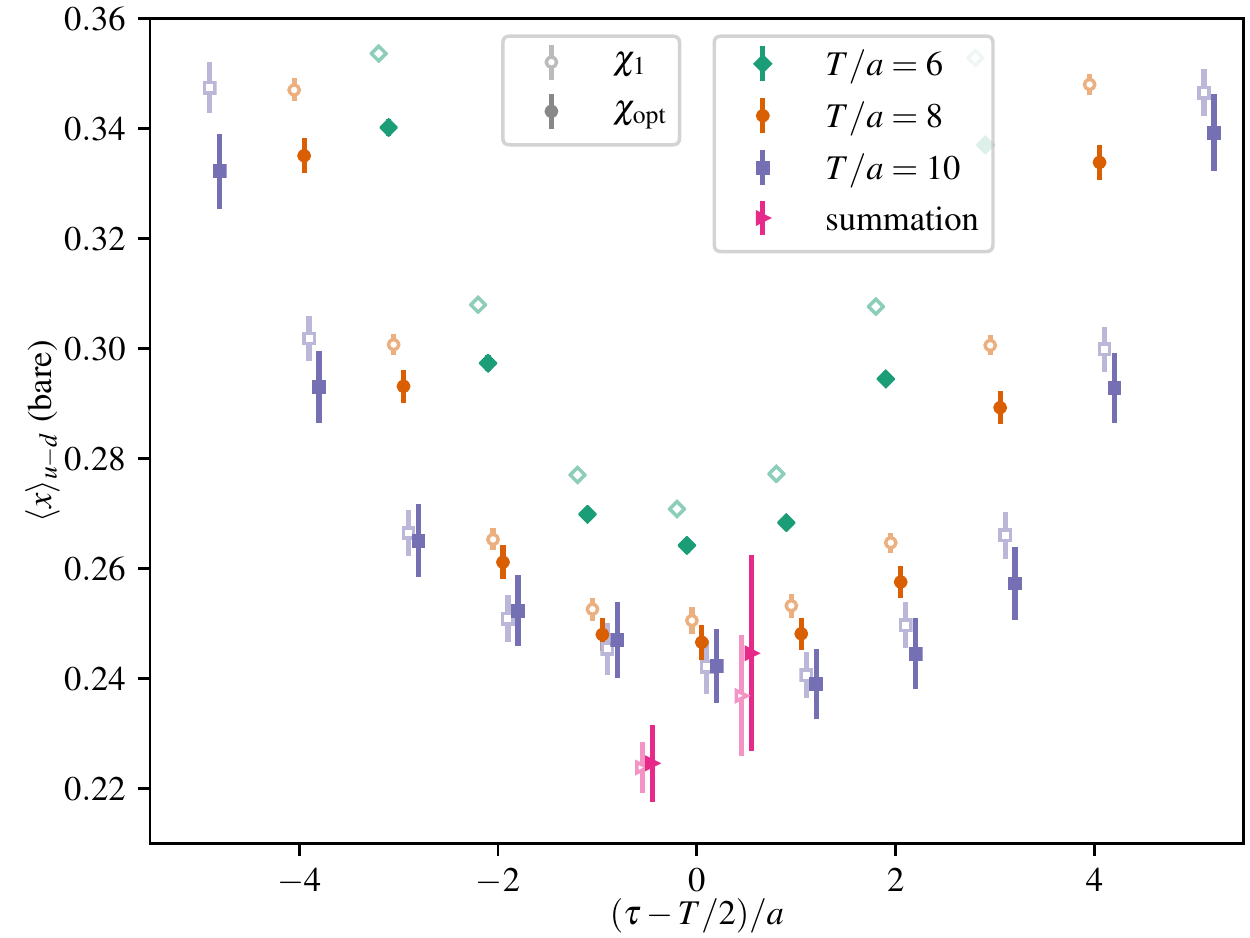}
  \caption{Isovector axial, tensor, and scalar charges ($g_A$, $g_T$,
    and $g_S$) and isovector momentum fraction
    $\langle x\rangle_{u-d}$ determined using the standard
    interpolator $\chi_1$ (open symbols) and the linear combination of
    standard and hybrid interpolators $\chi_\text{opt}$ (filled
    symbols). Ratio-method data are shown for source-sink separations
    $T/a=6$ (green diamonds), 8 (orange circles), and 10 (blue
    squares), and are plotted versus the operator insertion time
    $\tau$, shifted by half the source-sink
    separation. Summation-method (magenta triangles) data are based on
    the discrete derivative of sums, $[S(T+2a)-S(T)]/(2a)$; the two
    points correspond to $T/a=6$ and 8.}
  \label{fig:charges}
\end{figure*}

Results are shown in the ``plateau plots'' of
Fig.~\ref{fig:charges}. The behavior depends strongly on the
observable. For $g_A$, $\chi_\text{opt}$ produces a much stronger
dependence on the operator insertion time, $\tau$, indicating a
significant enhancement of excited-state contributions. The opposite
is true for $g_T$, where excited-state effects appear to be
significantly suppressed when using $\chi_\text{opt}$. The difference
between the two operators is relatively small for $g_S$ and
$\langle x\rangle_{u-d}$.
It is particularly problematic that
$\chi_\text{opt}$, which appears to be an improved operator based on
the two-point correlator, produces significantly enhanced
excited-state effects in $g_A$. In addition, across all observables
$\chi_\text{opt}$ produces consistently larger statistical
uncertainties.

We have also explored the use of simultaneous four-state fits to
two-point and three-point correlators. However, for each operator
$\mathcal{O}$ the corresponding fit model has ten independent unknown
matrix elements $\langle n'|\mathcal{O}|n\rangle$. Since we have
computed three-point correlators with only two combinations of source
and sink interpolators, the fits are unable to constrain most of the
operator matrix elements. This prevents the use of four-state fits to
understand what causes the differences between $\chi_1$ and
$\chi_\text{opt}$ for estimating matrix elements.

\subsection{Form factors}
\label{sec:form_factors}

\begin{figure*}
  \includegraphics[width=0.8\textwidth]{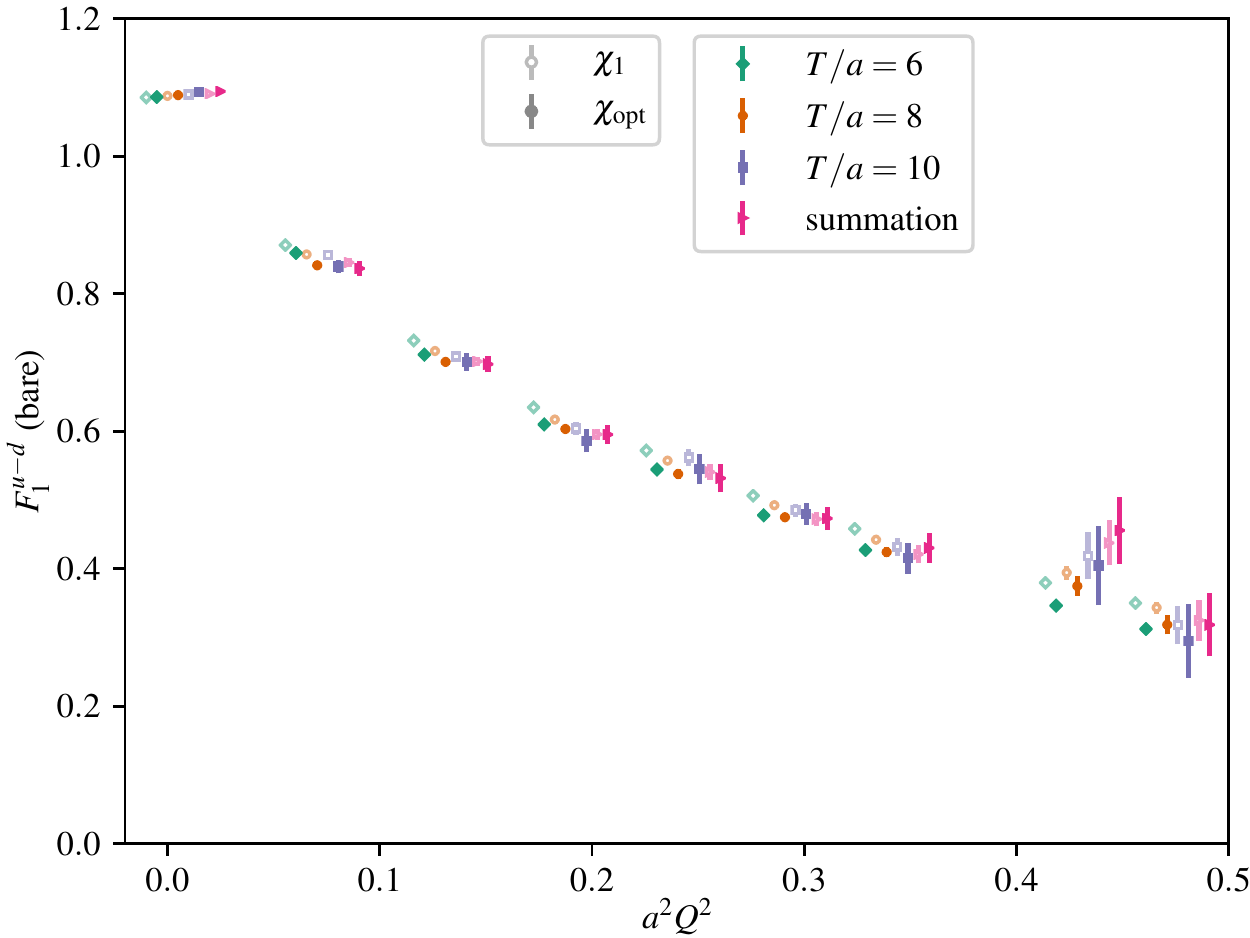}\\
  \includegraphics[width=0.8\textwidth]{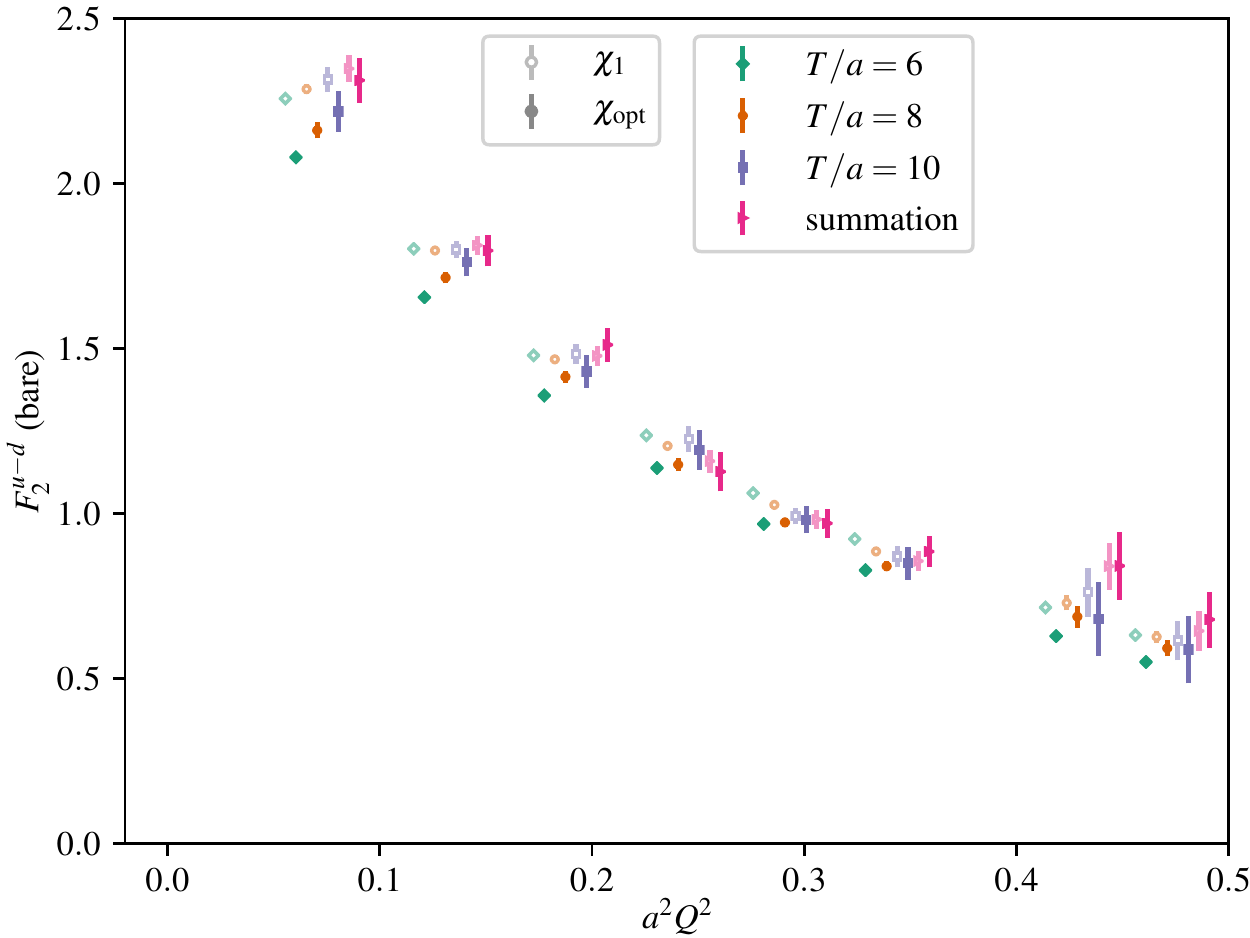}
  \caption{Isovector Dirac and Pauli form factors, $F_1$ and $F_2$,
    determined using the standard interpolator $\chi_1$ (open symbols)
    and the linear combination of standard and hybrid interpolators
    $\chi_\text{opt}$ (filled symbols). Ratio-method data are shown
    for source-sink separations $T/a=6$ (green diamonds), 8 (orange
    circles), and 10 (blue squares). Summation-method (magenta
    triangles) data are based on fitting a line to the three sums.}
  \label{fig:F1F2}
\end{figure*}

\begin{figure*}
  \includegraphics[width=0.8\textwidth]{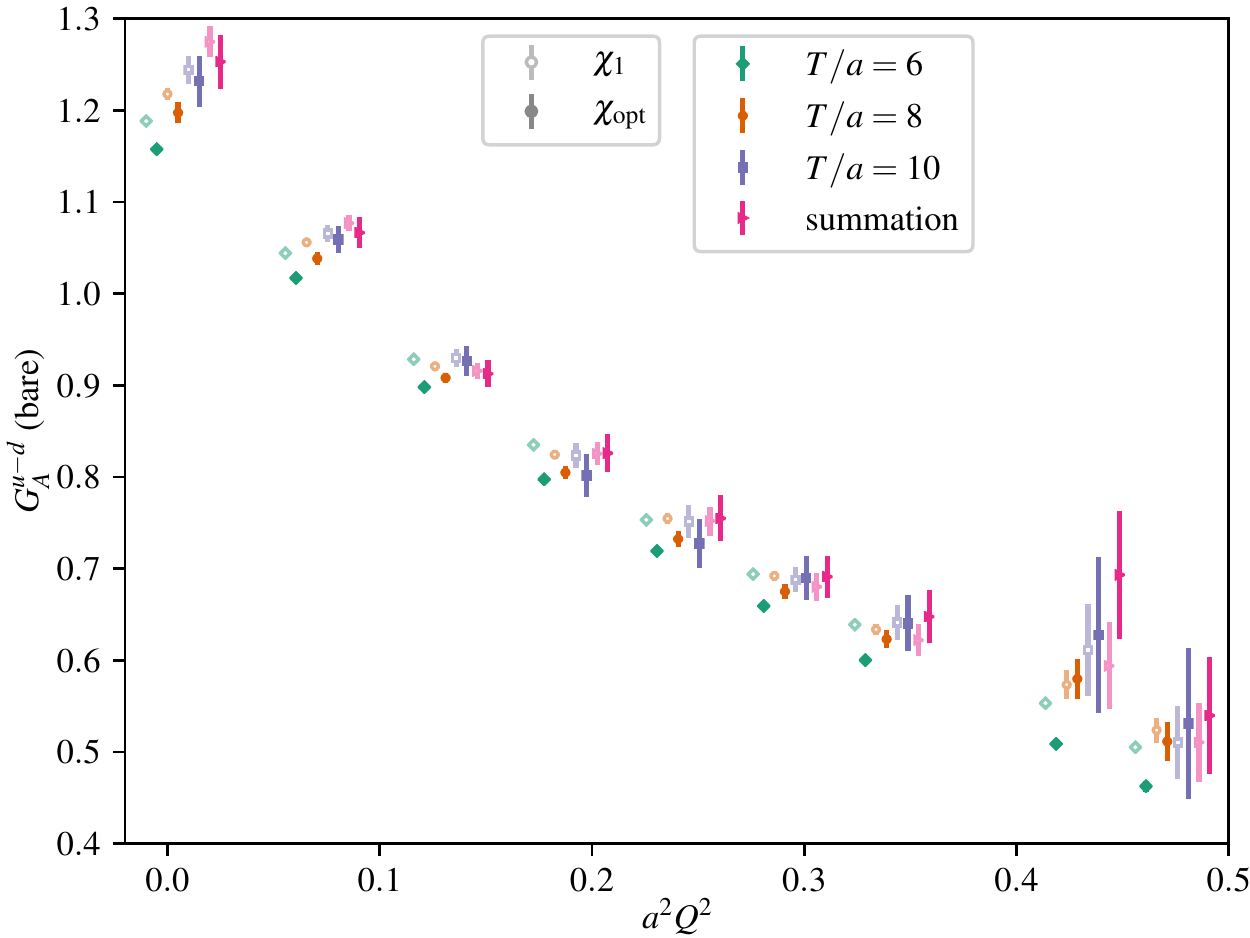}\\
  \includegraphics[width=0.8\textwidth]{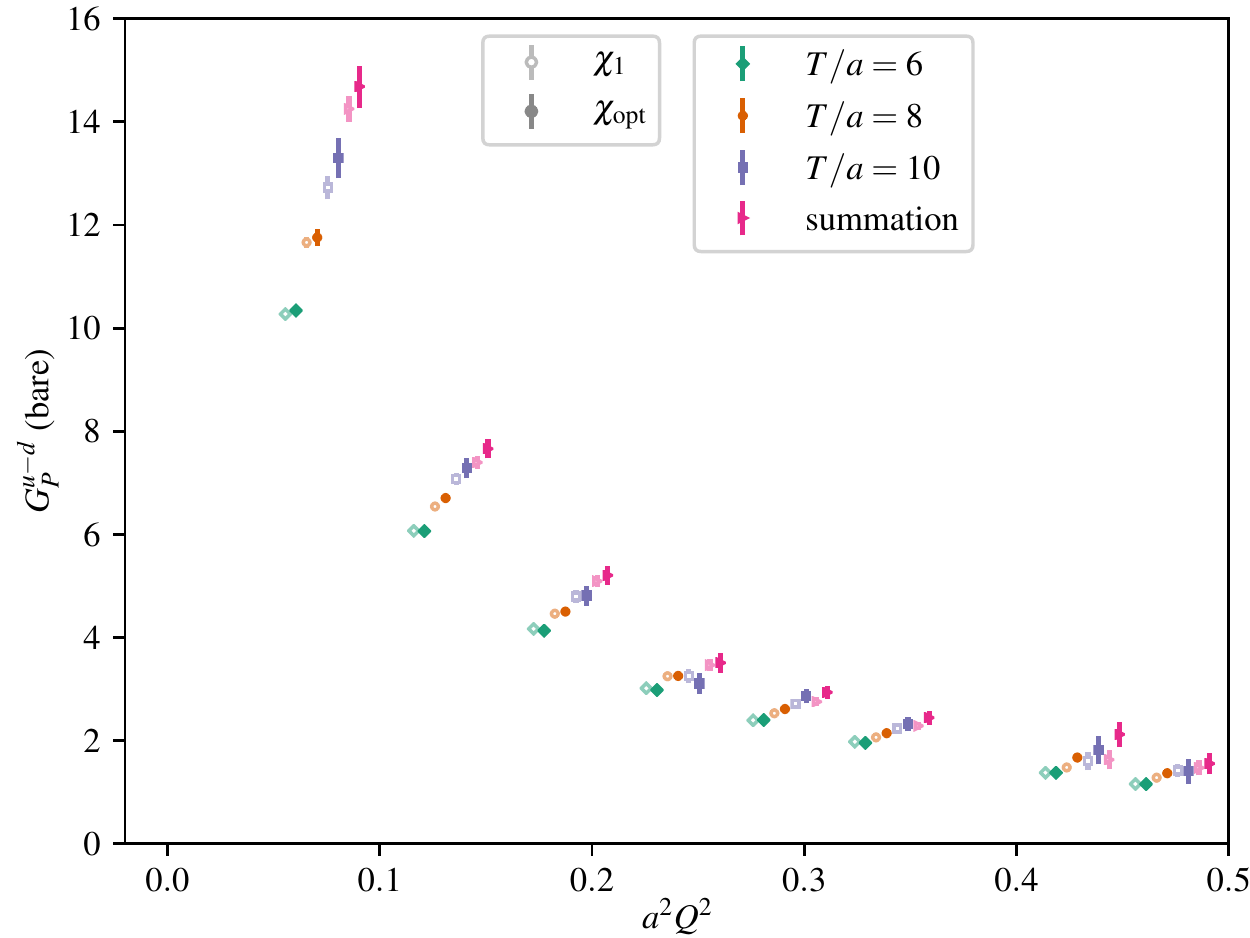}
  \caption{Isovector axial and induced pseudoscalar form factors,
    $G_A$ and $G_P$. See the caption of Fig.~\ref{fig:F1F2}.}
  \label{fig:GAGP}
\end{figure*}

The isovector form factors $F_1$ and $F_2$ of the electromagnetic
current are shown in Fig.~\ref{fig:F1F2} and the form factors $G_A$
and $G_P$ of the axial current are shown in Fig.~\ref{fig:GAGP}. The
results are rather mixed. For $F_2$ near $a^2Q^2=0.3$ and for $F_1$,
$\chi_\text{opt}$ produces a weaker dependence on the source-sink
separation than $\chi_1$, indicating the suppression of excited-state
contributions. However, the opposite is true for $F_2$ at lower $Q^2$
and for $G_A$. For $G_P$, the two operators produce similar
excited-state effects, which are very large at low $Q^2$. Chiral
perturbation theory predicts this large excited-state
contribution~\cite{Bar:2018xyi} as a result of nucleon-pion states,
which the hybrid operators are unlikely to help remove.

\section{Conclusions}
\label{sec:conclusions}

The use of a variational basis comprising a standard interpolating
operator and hybrid ones presents the possibility of reducing
excited-state contamination in nucleon structure calculations at low
computational cost. In two-point correlators, this is borne out, as
seen in Figs.~\ref{fig:meff} and \ref{fig:Eeff}. However, in nucleon
structure observables that depend on three-point correlators there is
no consistent result. The tensor charge shows significantly reduced
excited-state effects, whereas the axial charge shows increased
effects\footnote{The opposite effect was seen using a different basis
  of interpolators in Ref.~\cite{Yoon:2016dij}, where the variational
  analyses produced reduced excited-state effects for $g_A$ and
  enhanced excited-state effects for $g_T$, compared with the
  interpolator $S_5$.}. Other observables show little change and for
form factors the result can depend on $Q^2$. If one also takes into
account the increased statistical uncertainty, then the setup using
hybrid interpolators appears to be not worth pursuing further in its
current form.

Results from the four-state fit indicate
that the variational procedure succeeds at suppressing some excited
states while another lower-lying state is unaffected.
The presence of many relevant excited states suggests ways of
understanding the results: one possibility is that for the axial
charge, the variational procedure spoils a partial cancellation of
contributions between two different excited states. Since operator
matrix elements do not play a role in the GEVP, there is no guarantee
that reducing the net contribution from excited states in the
two-point correlator will do the same in three-point correlators.

Reliably improved results using the variational method can only be
obtained by being in the asymptotic regime of
Ref.~\cite{Blossier:2009kd}. This means that all low-lying excitations
must be identified, including multiparticle states, which in practice
require nonlocal interpolators. Doing so amounts to a challenging
computation; however, as this work has shown, half-measures (such as
using a small number of local interpolators) do not consistently
reduce excited-state contamination in nucleon structure observables.

\acknowledgments

We thank the Budapest-Marseille-Wuppertal collaboration for making
their configurations available to us. Calculations for this project
were done using the Qlua software suite~\cite{Qlua}, and made use of
the QOPQDP adaptive multigrid solver~\cite{Babich:2010qb,QOPQDP}.

This research used resources on the supercomputer JUWELS~\cite{juwels}
at Jülich Supercomputing Centre (JSC); we acknowledge computing time
granted by the John von Neumann Institute for Computing (NIC).

SM is supported by the U.S. Department of Energy (DOE), Office of
Science, Office of High Energy Physics under Award Number
DE-SC0009913. SM and SS are also supported by the RIKEN BNL Research
Center under its joint tenure track fellowships with the University of
Arizona and Stony Brook University, respectively. ME, JN, and AP are
supported in part by the Office of Nuclear Physics of the U.S. DOE
under grants DE-FG02-96ER40965, DE-SC-0011090, and DE-SC0018121,
respectively. SK and NH received support from Deutsche
Forschungsgemeinschaft grant SFB-TRR 55.

\bibliography{hybrid_operators}

\begin{thebibliography}{33}%
\makeatletter
\providecommand \@ifxundefined [1]{%
 \@ifx{#1\undefined}
}%
\providecommand \@ifnum [1]{%
 \ifnum #1\expandafter \@firstoftwo
 \else \expandafter \@secondoftwo
 \fi
}%
\providecommand \@ifx [1]{%
 \ifx #1\expandafter \@firstoftwo
 \else \expandafter \@secondoftwo
 \fi
}%
\providecommand \natexlab [1]{#1}%
\providecommand \enquote  [1]{``#1''}%
\providecommand \bibnamefont  [1]{#1}%
\providecommand \bibfnamefont [1]{#1}%
\providecommand \citenamefont [1]{#1}%
\providecommand \href@noop [0]{\@secondoftwo}%
\providecommand \href [0]{\begingroup \@sanitize@url \@href}%
\providecommand \@href[1]{\@@startlink{#1}\@@href}%
\providecommand \@@href[1]{\endgroup#1\@@endlink}%
\providecommand \@sanitize@url [0]{\catcode `\\12\catcode `\$12\catcode
  `\&12\catcode `\#12\catcode `\^12\catcode `\_12\catcode `\%12\relax}%
\providecommand \@@startlink[1]{}%
\providecommand \@@endlink[0]{}%
\providecommand \url  [0]{\begingroup\@sanitize@url \@url }%
\providecommand \@url [1]{\endgroup\@href {#1}{\urlprefix }}%
\providecommand \urlprefix  [0]{URL }%
\providecommand \Eprint [0]{\href }%
\providecommand \doibase [0]{http://dx.doi.org/}%
\providecommand \selectlanguage [0]{\@gobble}%
\providecommand \bibinfo  [0]{\@secondoftwo}%
\providecommand \bibfield  [0]{\@secondoftwo}%
\providecommand \translation [1]{[#1]}%
\providecommand \BibitemOpen [0]{}%
\providecommand \bibitemStop [0]{}%
\providecommand \bibitemNoStop [0]{.\EOS\space}%
\providecommand \EOS [0]{\spacefactor3000\relax}%
\providecommand \BibitemShut  [1]{\csname bibitem#1\endcsname}%
\let\auto@bib@innerbib\@empty
\bibitem [{\citenamefont {Lepage}(1989)}]{Lepage:1989hd}%
  \BibitemOpen
  \bibfield  {author} {\bibinfo {author} {\bibfnamefont {G.~P.}\ \bibnamefont
  {Lepage}},\ }\bibfield  {title} {\enquote {\bibinfo {title} {The analysis of
  algorithms for lattice field theory},}\ }in\ \href@noop {} {\emph {\bibinfo
  {booktitle} {From Actions to Answers: Proceedings of the 1989 {Theoretical
  Advanced Study Institute} in Elementary Particle Physics}}},\ \bibinfo
  {editor} {edited by\ \bibinfo {editor} {\bibfnamefont {T.}~\bibnamefont
  {DeGrand}}\ and\ \bibinfo {editor} {\bibfnamefont {D.}~\bibnamefont
  {Toussaint}}}\ (\bibinfo  {publisher} {World Scientific},\ \bibinfo {year}
  {1989})\ pp.\ \bibinfo {pages} {97--120},\ \bibinfo {note}
  {\url{http://inspirehep.net/record/287173}}\BibitemShut {NoStop}%
\bibitem [{\citenamefont {Michael}(1985)}]{Michael:1985ne}%
  \BibitemOpen
  \bibfield  {author} {\bibinfo {author} {\bibfnamefont {C.}~\bibnamefont
  {Michael}},\ }\bibfield  {title} {\enquote {\bibinfo {title} {Adjoint sources
  in lattice gauge theory},}\ }\href {\doibase 10.1016/0550-3213(85)90297-4}
  {\bibfield  {journal} {\bibinfo  {journal} {Nucl. Phys. B}\ }\textbf
  {\bibinfo {volume} {259}},\ \bibinfo {pages} {58} (\bibinfo {year}
  {1985})}\BibitemShut {NoStop}%
\bibitem [{\citenamefont {Lüscher}\ and\ \citenamefont
  {Wolff}(1990)}]{Luscher:1990ck}%
  \BibitemOpen
  \bibfield  {author} {\bibinfo {author} {\bibfnamefont {M.}~\bibnamefont
  {Lüscher}}\ and\ \bibinfo {author} {\bibfnamefont {U.}~\bibnamefont
  {Wolff}},\ }\bibfield  {title} {\enquote {\bibinfo {title} {How to calculate
  the elastic scattering matrix in two-dimensional quantum field theories by
  numerical simulation},}\ }\href {\doibase 10.1016/0550-3213(90)90540-T}
  {\bibfield  {journal} {\bibinfo  {journal} {Nucl. Phys. B}\ }\textbf
  {\bibinfo {volume} {339}},\ \bibinfo {pages} {222} (\bibinfo {year}
  {1990})}\BibitemShut {NoStop}%
\bibitem [{\citenamefont {Blossier}\ \emph {et~al.}(2009)\citenamefont
  {Blossier}, \citenamefont {Della~Morte}, \citenamefont {von Hippel},
  \citenamefont {Mendes},\ and\ \citenamefont {Sommer}}]{Blossier:2009kd}%
  \BibitemOpen
  \bibfield  {author} {\bibinfo {author} {\bibfnamefont {B.}~\bibnamefont
  {Blossier}}, \bibinfo {author} {\bibfnamefont {M.}~\bibnamefont
  {Della~Morte}}, \bibinfo {author} {\bibfnamefont {G.}~\bibnamefont {von
  Hippel}}, \bibinfo {author} {\bibfnamefont {T.}~\bibnamefont {Mendes}}, \
  and\ \bibinfo {author} {\bibfnamefont {R.}~\bibnamefont {Sommer}},\
  }\bibfield  {title} {\enquote {\bibinfo {title} {{On the generalized
  eigenvalue method for energies and matrix elements in lattice field
  theory}},}\ }\href {\doibase 10.1088/1126-6708/2009/04/094} {\bibfield
  {journal} {\bibinfo  {journal} {JHEP}\ }\textbf {\bibinfo {volume} {04}},\
  \bibinfo {pages} {094} (\bibinfo {year} {2009})},\ \Eprint
  {http://arxiv.org/abs/0902.1265} {arXiv:0902.1265 [hep-lat]} \BibitemShut
  {NoStop}%
\bibitem [{\citenamefont {Engel}\ \emph {et~al.}(2009)\citenamefont {Engel},
  \citenamefont {Gattringer}, \citenamefont {Glozman}, \citenamefont {Lang},
  \citenamefont {Limmer}, \citenamefont {Mohler},\ and\ \citenamefont
  {Schäfer}}]{Engel:2009nh}%
  \BibitemOpen
  \bibfield  {author} {\bibinfo {author} {\bibfnamefont {G.}~\bibnamefont
  {Engel}}, \bibinfo {author} {\bibfnamefont {C.}~\bibnamefont {Gattringer}},
  \bibinfo {author} {\bibfnamefont {L.~{\relax Ya}.}\ \bibnamefont {Glozman}},
  \bibinfo {author} {\bibfnamefont {C.~B.}\ \bibnamefont {Lang}}, \bibinfo
  {author} {\bibfnamefont {M.}~\bibnamefont {Limmer}}, \bibinfo {author}
  {\bibfnamefont {D.}~\bibnamefont {Mohler}}, \ and\ \bibinfo {author}
  {\bibfnamefont {A.}~\bibnamefont {Schäfer}},\ }\bibfield  {title} {\enquote
  {\bibinfo {title} {{Baryon axial charges from Chirally Improved fermions:
  First results}},}\ }\bibfield  {booktitle} {\emph {\bibinfo {booktitle}
  {{Proceedings, 27th International Symposium on Lattice field theory (Lattice
  2009): Beijing, P.R. China, July 26-31, 2009}}},\ }\href {\doibase
  10.22323/1.091.0135} {\bibfield  {journal} {\bibinfo  {journal} {PoS}\
  }\textbf {\bibinfo {volume} {LAT2009}},\ \bibinfo {pages} {135} (\bibinfo
  {year} {2009})},\ \Eprint {http://arxiv.org/abs/0910.4190} {arXiv:0910.4190
  [hep-lat]} \BibitemShut {NoStop}%
\bibitem [{\citenamefont {Owen}\ \emph {et~al.}(2013)\citenamefont {Owen},
  \citenamefont {Dragos}, \citenamefont {Kamleh}, \citenamefont {Leinweber},
  \citenamefont {Mahbub}, \citenamefont {Menadue},\ and\ \citenamefont
  {Zanotti}}]{Owen:2012ts}%
  \BibitemOpen
  \bibfield  {author} {\bibinfo {author} {\bibfnamefont {B.~J.}\ \bibnamefont
  {Owen}}, \bibinfo {author} {\bibfnamefont {J.}~\bibnamefont {Dragos}},
  \bibinfo {author} {\bibfnamefont {W.}~\bibnamefont {Kamleh}}, \bibinfo
  {author} {\bibfnamefont {D.~B.}\ \bibnamefont {Leinweber}}, \bibinfo {author}
  {\bibfnamefont {M.~S.}\ \bibnamefont {Mahbub}}, \bibinfo {author}
  {\bibfnamefont {B.~J.}\ \bibnamefont {Menadue}}, \ and\ \bibinfo {author}
  {\bibfnamefont {J.~M.}\ \bibnamefont {Zanotti}},\ }\bibfield  {title}
  {\enquote {\bibinfo {title} {Variational approach to the calculation of
  {$g_A$}},}\ }\href {\doibase 10.1016/j.physletb.2013.04.063} {\bibfield
  {journal} {\bibinfo  {journal} {Phys. Lett. B}\ }\textbf {\bibinfo {volume}
  {723}},\ \bibinfo {pages} {217} (\bibinfo {year} {2013})},\ \Eprint
  {http://arxiv.org/abs/1212.4668} {arXiv:1212.4668 [hep-lat]} \BibitemShut
  {NoStop}%
\bibitem [{\citenamefont {Yoon}\ \emph {et~al.}(2016)\citenamefont {Yoon} \emph
  {et~al.}}]{Yoon:2016dij}%
  \BibitemOpen
  \bibfield  {author} {\bibinfo {author} {\bibfnamefont {B.}~\bibnamefont
  {Yoon}} \emph {et~al.},\ }\bibfield  {title} {\enquote {\bibinfo {title}
  {Controlling excited-state contamination in nucleon matrix elements},}\
  }\href {\doibase 10.1103/PhysRevD.93.114506} {\bibfield  {journal} {\bibinfo
  {journal} {Phys. Rev. D}\ }\textbf {\bibinfo {volume} {93}},\ \bibinfo
  {pages} {114506} (\bibinfo {year} {2016})},\ \Eprint
  {http://arxiv.org/abs/1602.07737} {arXiv:1602.07737 [hep-lat]} \BibitemShut
  {NoStop}%
\bibitem [{\citenamefont {Dragos}\ \emph {et~al.}(2016)\citenamefont {Dragos},
  \citenamefont {Horsley}, \citenamefont {Kamleh}, \citenamefont {Leinweber},
  \citenamefont {Nakamura}, \citenamefont {Rakow}, \citenamefont {Schierholz},
  \citenamefont {Young},\ and\ \citenamefont {Zanotti}}]{Dragos:2016rtx}%
  \BibitemOpen
  \bibfield  {author} {\bibinfo {author} {\bibfnamefont {J.}~\bibnamefont
  {Dragos}}, \bibinfo {author} {\bibfnamefont {R.}~\bibnamefont {Horsley}},
  \bibinfo {author} {\bibfnamefont {W.}~\bibnamefont {Kamleh}}, \bibinfo
  {author} {\bibfnamefont {D.~B.}\ \bibnamefont {Leinweber}}, \bibinfo {author}
  {\bibfnamefont {Y.}~\bibnamefont {Nakamura}}, \bibinfo {author}
  {\bibfnamefont {P.~E.~L.}\ \bibnamefont {Rakow}}, \bibinfo {author}
  {\bibfnamefont {G.}~\bibnamefont {Schierholz}}, \bibinfo {author}
  {\bibfnamefont {R.~D.}\ \bibnamefont {Young}}, \ and\ \bibinfo {author}
  {\bibfnamefont {J.~M.}\ \bibnamefont {Zanotti}},\ }\bibfield  {title}
  {\enquote {\bibinfo {title} {{Nucleon matrix elements using the variational
  method in lattice QCD}},}\ }\href {\doibase 10.1103/PhysRevD.94.074505}
  {\bibfield  {journal} {\bibinfo  {journal} {Phys. Rev. D}\ }\textbf {\bibinfo
  {volume} {94}},\ \bibinfo {pages} {074505} (\bibinfo {year} {2016})},\
  \Eprint {http://arxiv.org/abs/1606.03195} {arXiv:1606.03195 [hep-lat]}
  \BibitemShut {NoStop}%
\bibitem [{\citenamefont {Stokes}\ \emph {et~al.}(2019)\citenamefont {Stokes},
  \citenamefont {Kamleh},\ and\ \citenamefont {Leinweber}}]{Stokes:2018emx}%
  \BibitemOpen
  \bibfield  {author} {\bibinfo {author} {\bibfnamefont {F.~M.}\ \bibnamefont
  {Stokes}}, \bibinfo {author} {\bibfnamefont {W.}~\bibnamefont {Kamleh}}, \
  and\ \bibinfo {author} {\bibfnamefont {D.~B.}\ \bibnamefont {Leinweber}},\
  }\bibfield  {title} {\enquote {\bibinfo {title} {Opposite-parity
  contaminations in lattice nucleon form factors},}\ }\href {\doibase
  10.1103/PhysRevD.99.074506} {\bibfield  {journal} {\bibinfo  {journal} {Phys.
  Rev. D}\ }\textbf {\bibinfo {volume} {99}},\ \bibinfo {pages} {074506}
  (\bibinfo {year} {2019})},\ \Eprint {http://arxiv.org/abs/1809.11002}
  {arXiv:1809.11002 [hep-lat]} \BibitemShut {NoStop}%
\bibitem [{\citenamefont {Egerer}\ \emph {et~al.}(2019)\citenamefont {Egerer},
  \citenamefont {Richards},\ and\ \citenamefont {Winter}}]{Egerer:2018xgu}%
  \BibitemOpen
  \bibfield  {author} {\bibinfo {author} {\bibfnamefont {C.}~\bibnamefont
  {Egerer}}, \bibinfo {author} {\bibfnamefont {D.}~\bibnamefont {Richards}}, \
  and\ \bibinfo {author} {\bibfnamefont {F.}~\bibnamefont {Winter}},\
  }\bibfield  {title} {\enquote {\bibinfo {title} {{Controlling excited-state
  contributions with distillation in lattice QCD calculations of nucleon
  isovector charges $g_S^{u-d}$, $g_A^{u-d}$, $g_T^{u-d}$}},}\ }\href {\doibase
  10.1103/PhysRevD.99.034506} {\bibfield  {journal} {\bibinfo  {journal} {Phys.
  Rev. D}\ }\textbf {\bibinfo {volume} {99}},\ \bibinfo {pages} {034506}
  (\bibinfo {year} {2019})},\ \Eprint {http://arxiv.org/abs/1810.09991}
  {arXiv:1810.09991 [hep-lat]} \BibitemShut {NoStop}%
\bibitem [{\citenamefont {Dudek}\ and\ \citenamefont
  {Edwards}(2012)}]{Dudek:2012ag}%
  \BibitemOpen
  \bibfield  {author} {\bibinfo {author} {\bibfnamefont {J.~J.}\ \bibnamefont
  {Dudek}}\ and\ \bibinfo {author} {\bibfnamefont {R.~G.}\ \bibnamefont
  {Edwards}},\ }\bibfield  {title} {\enquote {\bibinfo {title} {{Hybrid Baryons
  in QCD}},}\ }\href {\doibase 10.1103/PhysRevD.85.054016} {\bibfield
  {journal} {\bibinfo  {journal} {Phys. Rev. D}\ }\textbf {\bibinfo {volume}
  {85}},\ \bibinfo {pages} {054016} (\bibinfo {year} {2012})},\ \Eprint
  {http://arxiv.org/abs/1201.2349} {arXiv:1201.2349 [hep-ph]} \BibitemShut
  {NoStop}%
\bibitem [{\citenamefont {Dürr}\ \emph {et~al.}(2011)\citenamefont {Dürr},
  \citenamefont {Fodor}, \citenamefont {Hoelbling}, \citenamefont {Katz},
  \citenamefont {Krieg}, \citenamefont {Kurth}, \citenamefont {Lellouch},
  \citenamefont {Lippert}, \citenamefont {Szabó},\ and\ \citenamefont
  {Vulvert}}]{Durr:2010aw}%
  \BibitemOpen
  \bibfield  {author} {\bibinfo {author} {\bibfnamefont {S.}~\bibnamefont
  {Dürr}}, \bibinfo {author} {\bibfnamefont {Z.}~\bibnamefont {Fodor}},
  \bibinfo {author} {\bibfnamefont {C.}~\bibnamefont {Hoelbling}}, \bibinfo
  {author} {\bibfnamefont {S.~D.}\ \bibnamefont {Katz}}, \bibinfo {author}
  {\bibfnamefont {S.}~\bibnamefont {Krieg}}, \bibinfo {author} {\bibfnamefont
  {T.}~\bibnamefont {Kurth}}, \bibinfo {author} {\bibfnamefont
  {L.}~\bibnamefont {Lellouch}}, \bibinfo {author} {\bibfnamefont
  {T.}~\bibnamefont {Lippert}}, \bibinfo {author} {\bibfnamefont {K.~K.}\
  \bibnamefont {Szabó}}, \ and\ \bibinfo {author} {\bibfnamefont
  {G.}~\bibnamefont {Vulvert}},\ }\bibfield  {title} {\enquote {\bibinfo
  {title} {{Lattice QCD at the physical point: Simulation and analysis
  details}},}\ }\href {\doibase 10.1007/JHEP08(2011)148} {\bibfield  {journal}
  {\bibinfo  {journal} {JHEP}\ }\textbf {\bibinfo {volume} {08}},\ \bibinfo
  {pages} {148} (\bibinfo {year} {2011})},\ \Eprint
  {http://arxiv.org/abs/1011.2711} {arXiv:1011.2711 [hep-lat]} \BibitemShut
  {NoStop}%
\bibitem [{\citenamefont {Green}\ \emph
  {et~al.}(2014{\natexlab{a}})\citenamefont {Green}, \citenamefont {Negele},
  \citenamefont {Pochinsky}, \citenamefont {Syritsyn}, \citenamefont
  {Engelhardt},\ and\ \citenamefont {Krieg}}]{Green:2014xba}%
  \BibitemOpen
  \bibfield  {author} {\bibinfo {author} {\bibfnamefont {J.~R.}\ \bibnamefont
  {Green}}, \bibinfo {author} {\bibfnamefont {J.~W.}\ \bibnamefont {Negele}},
  \bibinfo {author} {\bibfnamefont {A.~V.}\ \bibnamefont {Pochinsky}}, \bibinfo
  {author} {\bibfnamefont {S.~N.}\ \bibnamefont {Syritsyn}}, \bibinfo {author}
  {\bibfnamefont {M.}~\bibnamefont {Engelhardt}}, \ and\ \bibinfo {author}
  {\bibfnamefont {S.}~\bibnamefont {Krieg}},\ }\bibfield  {title} {\enquote
  {\bibinfo {title} {{Nucleon electromagnetic form factors from lattice QCD
  using a nearly physical pion mass}},}\ }\href {\doibase
  10.1103/PhysRevD.90.074507} {\bibfield  {journal} {\bibinfo  {journal} {Phys.
  Rev. D}\ }\textbf {\bibinfo {volume} {90}},\ \bibinfo {pages} {074507}
  (\bibinfo {year} {2014}{\natexlab{a}})},\ \Eprint
  {http://arxiv.org/abs/1404.4029} {arXiv:1404.4029 [hep-lat]} \BibitemShut
  {NoStop}%
\bibitem [{\citenamefont {Tiburzi}(2009)}]{Tiburzi:2009zp}%
  \BibitemOpen
  \bibfield  {author} {\bibinfo {author} {\bibfnamefont {B.~C.}\ \bibnamefont
  {Tiburzi}},\ }\bibfield  {title} {\enquote {\bibinfo {title} {Time dependence
  of nucleon correlation functions in chiral perturbation theory},}\ }\href
  {\doibase 10.1103/PhysRevD.80.014002} {\bibfield  {journal} {\bibinfo
  {journal} {Phys. Rev. D}\ }\textbf {\bibinfo {volume} {80}},\ \bibinfo
  {pages} {014002} (\bibinfo {year} {2009})},\ \Eprint
  {http://arxiv.org/abs/0901.0657} {arXiv:0901.0657 [hep-lat]} \BibitemShut
  {NoStop}%
\bibitem [{\citenamefont {Tiburzi}(2015)}]{Tiburzi:2015tta}%
  \BibitemOpen
  \bibfield  {author} {\bibinfo {author} {\bibfnamefont {B.~C.}\ \bibnamefont
  {Tiburzi}},\ }\bibfield  {title} {\enquote {\bibinfo {title} {Chiral
  corrections to nucleon two- and three-point correlation functions},}\ }\href
  {\doibase 10.1103/PhysRevD.91.094510} {\bibfield  {journal} {\bibinfo
  {journal} {Phys. Rev. D}\ }\textbf {\bibinfo {volume} {91}},\ \bibinfo
  {pages} {094510} (\bibinfo {year} {2015})},\ \Eprint
  {http://arxiv.org/abs/1503.06329} {arXiv:1503.06329 [hep-lat]} \BibitemShut
  {NoStop}%
\bibitem [{\citenamefont {Bär}(2015)}]{Bar:2015zwa}%
  \BibitemOpen
  \bibfield  {author} {\bibinfo {author} {\bibfnamefont {O.}~\bibnamefont
  {Bär}},\ }\bibfield  {title} {\enquote {\bibinfo {title}
  {{Nucleon-pion-state contribution to nucleon two-point correlation
  functions}},}\ }\href {\doibase 10.1103/PhysRevD.92.074504} {\bibfield
  {journal} {\bibinfo  {journal} {Phys. Rev. D}\ }\textbf {\bibinfo {volume}
  {92}},\ \bibinfo {pages} {074504} (\bibinfo {year} {2015})},\ \Eprint
  {http://arxiv.org/abs/1503.03649} {arXiv:1503.03649 [hep-lat]} \BibitemShut
  {NoStop}%
\bibitem [{\citenamefont {Bär}(2016)}]{Bar:2016uoj}%
  \BibitemOpen
  \bibfield  {author} {\bibinfo {author} {\bibfnamefont {O.}~\bibnamefont
  {Bär}},\ }\bibfield  {title} {\enquote {\bibinfo {title}
  {{Nucleon-pion-state contribution in lattice calculations of the nucleon
  charges $g_A,g_T$ and $g_S$}},}\ }\href {\doibase 10.1103/PhysRevD.94.054505}
  {\bibfield  {journal} {\bibinfo  {journal} {Phys. Rev. D}\ }\textbf {\bibinfo
  {volume} {94}},\ \bibinfo {pages} {054505} (\bibinfo {year} {2016})},\
  \Eprint {http://arxiv.org/abs/1606.09385} {arXiv:1606.09385 [hep-lat]}
  \BibitemShut {NoStop}%
\bibitem [{\citenamefont {Hansen}\ and\ \citenamefont
  {Meyer}(2017)}]{Hansen:2016qoz}%
  \BibitemOpen
  \bibfield  {author} {\bibinfo {author} {\bibfnamefont {M.~T.}\ \bibnamefont
  {Hansen}}\ and\ \bibinfo {author} {\bibfnamefont {H.~B.}\ \bibnamefont
  {Meyer}},\ }\bibfield  {title} {\enquote {\bibinfo {title} {{On the effect of
  excited states in lattice calculations of the nucleon axial charge}},}\
  }\href {\doibase 10.1016/j.nuclphysb.2017.08.017} {\bibfield  {journal}
  {\bibinfo  {journal} {Nucl. Phys. B}\ }\textbf {\bibinfo {volume} {923}},\
  \bibinfo {pages} {558} (\bibinfo {year} {2017})},\ \Eprint
  {http://arxiv.org/abs/1610.03843} {arXiv:1610.03843 [hep-lat]} \BibitemShut
  {NoStop}%
\bibitem [{\citenamefont {Bär}(2017)}]{Bar:2017kxh}%
  \BibitemOpen
  \bibfield  {author} {\bibinfo {author} {\bibfnamefont {O.}~\bibnamefont
  {Bär}},\ }\bibfield  {title} {\enquote {\bibinfo {title} {{Chiral
  perturbation theory and nucleon–pion-state contaminations in lattice
  QCD}},}\ }\href {\doibase 10.1142/S0217751X17300113} {\bibfield  {journal}
  {\bibinfo  {journal} {Int. J. Mod. Phys. A}\ }\textbf {\bibinfo {volume}
  {32}},\ \bibinfo {pages} {1730011} (\bibinfo {year} {2017})},\ \Eprint
  {http://arxiv.org/abs/1705.02806} {arXiv:1705.02806 [hep-lat]} \BibitemShut
  {NoStop}%
\bibitem [{\citenamefont {Bär}(2019)}]{Bar:2018xyi}%
  \BibitemOpen
  \bibfield  {author} {\bibinfo {author} {\bibfnamefont {O.}~\bibnamefont
  {Bär}},\ }\bibfield  {title} {\enquote {\bibinfo {title} {{$N\pi$-state
  contamination in lattice calculations of the nucleon axial form factors}},}\
  }\href {\doibase 10.1103/PhysRevD.99.054506} {\bibfield  {journal} {\bibinfo
  {journal} {Phys. Rev. D}\ }\textbf {\bibinfo {volume} {99}},\ \bibinfo
  {pages} {054506} (\bibinfo {year} {2019})},\ \Eprint
  {http://arxiv.org/abs/1812.09191} {arXiv:1812.09191 [hep-lat]} \BibitemShut
  {NoStop}%
\bibitem [{\citenamefont {Green}(2018)}]{Green:2018vxw}%
  \BibitemOpen
  \bibfield  {author} {\bibinfo {author} {\bibfnamefont {J.}~\bibnamefont
  {Green}},\ }\bibfield  {title} {\enquote {\bibinfo {title} {{Systematics in
  nucleon matrix element calculations}},}\ }\bibfield  {booktitle} {\emph
  {\bibinfo {booktitle} {{Proceedings, 36th International Symposium on Lattice
  Field Theory (Lattice 2018): East Lansing, MI, United States, July 22-28,
  2018}}},\ }\href {\doibase 10.22323/1.334.0016} {\bibfield  {journal}
  {\bibinfo  {journal} {PoS}\ }\textbf {\bibinfo {volume} {LATTICE2018}},\
  \bibinfo {pages} {016} (\bibinfo {year} {2018})},\ \Eprint
  {http://arxiv.org/abs/1812.10574} {arXiv:1812.10574 [hep-lat]} \BibitemShut
  {NoStop}%
\bibitem [{\citenamefont {Dudek}\ \emph {et~al.}(2013)\citenamefont {Dudek},
  \citenamefont {Edwards},\ and\ \citenamefont {Thomas}}]{Dudek:2012xn}%
  \BibitemOpen
  \bibfield  {author} {\bibinfo {author} {\bibfnamefont {J.~J.}\ \bibnamefont
  {Dudek}}, \bibinfo {author} {\bibfnamefont {R.~G.}\ \bibnamefont {Edwards}},
  \ and\ \bibinfo {author} {\bibfnamefont {C.~E.}\ \bibnamefont {Thomas}}
  (\bibinfo {collaboration} {Hadron Spectrum}),\ }\bibfield  {title} {\enquote
  {\bibinfo {title} {{Energy dependence of the $\rho$ resonance in $\pi\pi$
  elastic scattering from lattice QCD}},}\ }\href {\doibase
  10.1103/PhysRevD.87.034505} {\bibfield  {journal} {\bibinfo  {journal} {Phys.
  Rev. D}\ }\textbf {\bibinfo {volume} {87}},\ \bibinfo {pages} {034505}
  (\bibinfo {year} {2013})},\ \bibinfo {note} {[Erratum:
  \href{http://dx.doi.org/10.1103/PhysRevD.90.099902}{Phys. Rev. D \textbf{90},
  099902 (2014)}]},\ \Eprint {http://arxiv.org/abs/1212.0830} {arXiv:1212.0830
  [hep-ph]} \BibitemShut {NoStop}%
\bibitem [{\citenamefont {Göckeler}\ \emph {et~al.}(1996)\citenamefont
  {Göckeler}, \citenamefont {Horsley}, \citenamefont {Ilgenfritz},
  \citenamefont {Perlt}, \citenamefont {Rakow}, \citenamefont {Schierholz},\
  and\ \citenamefont {Schiller}}]{Gockeler:1995wg}%
  \BibitemOpen
  \bibfield  {author} {\bibinfo {author} {\bibfnamefont {M.}~\bibnamefont
  {Göckeler}}, \bibinfo {author} {\bibfnamefont {R.}~\bibnamefont {Horsley}},
  \bibinfo {author} {\bibfnamefont {E.-M.}\ \bibnamefont {Ilgenfritz}},
  \bibinfo {author} {\bibfnamefont {H.}~\bibnamefont {Perlt}}, \bibinfo
  {author} {\bibfnamefont {P.~E.~L.}\ \bibnamefont {Rakow}}, \bibinfo {author}
  {\bibfnamefont {G.}~\bibnamefont {Schierholz}}, \ and\ \bibinfo {author}
  {\bibfnamefont {A.}~\bibnamefont {Schiller}},\ }\bibfield  {title} {\enquote
  {\bibinfo {title} {{Polarized and unpolarized nucleon structure functions
  from lattice QCD}},}\ }\href {\doibase 10.1103/PhysRevD.53.2317} {\bibfield
  {journal} {\bibinfo  {journal} {Phys. Rev. D}\ }\textbf {\bibinfo {volume}
  {53}},\ \bibinfo {pages} {2317} (\bibinfo {year} {1996})},\ \Eprint
  {http://arxiv.org/abs/hep-lat/9508004} {arXiv:hep-lat/9508004} \BibitemShut
  {NoStop}%
\bibitem [{\citenamefont {Basak}\ \emph {et~al.}(2005)\citenamefont {Basak},
  \citenamefont {Edwards}, \citenamefont {Fleming}, \citenamefont {Heller},
  \citenamefont {Morningstar}, \citenamefont {Richards}, \citenamefont {Sato},\
  and\ \citenamefont {Wallace}}]{Basak:2005ir}%
  \BibitemOpen
  \bibfield  {author} {\bibinfo {author} {\bibfnamefont {S.}~\bibnamefont
  {Basak}}, \bibinfo {author} {\bibfnamefont {R.}~\bibnamefont {Edwards}},
  \bibinfo {author} {\bibfnamefont {G.~T.}\ \bibnamefont {Fleming}}, \bibinfo
  {author} {\bibfnamefont {U.~M.}\ \bibnamefont {Heller}}, \bibinfo {author}
  {\bibfnamefont {C.}~\bibnamefont {Morningstar}}, \bibinfo {author}
  {\bibfnamefont {D.}~\bibnamefont {Richards}}, \bibinfo {author}
  {\bibfnamefont {I.}~\bibnamefont {Sato}}, \ and\ \bibinfo {author}
  {\bibfnamefont {S.~J.}\ \bibnamefont {Wallace}} (\bibinfo {collaboration}
  {Lattice Hadron Physics (LHPC)}),\ }\bibfield  {title} {\enquote {\bibinfo
  {title} {{Clebsch-Gordan construction of lattice interpolating fields for
  excited baryons}},}\ }\href {\doibase 10.1103/PhysRevD.72.074501} {\bibfield
  {journal} {\bibinfo  {journal} {Phys. Rev. D}\ }\textbf {\bibinfo {volume}
  {72}},\ \bibinfo {pages} {074501} (\bibinfo {year} {2005})},\ \Eprint
  {http://arxiv.org/abs/hep-lat/0508018} {arXiv:hep-lat/0508018} \BibitemShut
  {NoStop}%
\bibitem [{\citenamefont {Sheikholeslami}\ and\ \citenamefont
  {Wohlert}(1985)}]{Sheikholeslami:1985ij}%
  \BibitemOpen
  \bibfield  {author} {\bibinfo {author} {\bibfnamefont {B.}~\bibnamefont
  {Sheikholeslami}}\ and\ \bibinfo {author} {\bibfnamefont {R.}~\bibnamefont
  {Wohlert}},\ }\bibfield  {title} {\enquote {\bibinfo {title} {Improved
  continuum limit lattice action for {QCD} with {Wilson} fermions},}\ }\href
  {\doibase 10.1016/0550-3213(85)90002-1} {\bibfield  {journal} {\bibinfo
  {journal} {Nucl. Phys. B}\ }\textbf {\bibinfo {volume} {259}},\ \bibinfo
  {pages} {572} (\bibinfo {year} {1985})}\BibitemShut {NoStop}%
\bibitem [{\citenamefont {Güsken}(1990)}]{Gusken:1989qx}%
  \BibitemOpen
  \bibfield  {author} {\bibinfo {author} {\bibfnamefont {S.}~\bibnamefont
  {Güsken}},\ }\bibfield  {title} {\enquote {\bibinfo {title} {{A Study of
  smearing techniques for hadron correlation functions}},}\ }\bibfield
  {booktitle} {\emph {\bibinfo {booktitle} {{Lattice 89 Capri: The 1989
  Symposium on Lattice Field Theory Capri, Italy, September 18-21, 1989}}},\
  }\href {\doibase 10.1016/0920-5632(90)90273-W} {\bibfield  {journal}
  {\bibinfo  {journal} {Nucl. Phys. B (Proc. Suppl.)}\ }\textbf {\bibinfo
  {volume} {17}},\ \bibinfo {pages} {361} (\bibinfo {year} {1990})}\BibitemShut
  {NoStop}%
\bibitem [{\citenamefont {Green}\ \emph {et~al.}(2012)\citenamefont {Green},
  \citenamefont {Negele}, \citenamefont {Pochinsky}, \citenamefont {Syritsyn},
  \citenamefont {Engelhardt},\ and\ \citenamefont {Krieg}}]{Green:2012ej}%
  \BibitemOpen
  \bibfield  {author} {\bibinfo {author} {\bibfnamefont {J.~R.}\ \bibnamefont
  {Green}}, \bibinfo {author} {\bibfnamefont {J.~W.}\ \bibnamefont {Negele}},
  \bibinfo {author} {\bibfnamefont {A.~V.}\ \bibnamefont {Pochinsky}}, \bibinfo
  {author} {\bibfnamefont {S.~N.}\ \bibnamefont {Syritsyn}}, \bibinfo {author}
  {\bibfnamefont {M.}~\bibnamefont {Engelhardt}}, \ and\ \bibinfo {author}
  {\bibfnamefont {S.}~\bibnamefont {Krieg}},\ }\bibfield  {title} {\enquote
  {\bibinfo {title} {Nucleon scalar and tensor charges from lattice {QCD} with
  light {Wilson} quarks},}\ }\href {\doibase 10.1103/PhysRevD.86.114509}
  {\bibfield  {journal} {\bibinfo  {journal} {Phys. Rev. D}\ }\textbf {\bibinfo
  {volume} {86}},\ \bibinfo {pages} {114509} (\bibinfo {year} {2012})},\
  \Eprint {http://arxiv.org/abs/1206.4527} {arXiv:1206.4527 [hep-lat]}
  \BibitemShut {NoStop}%
\bibitem [{\citenamefont {Green}\ \emph
  {et~al.}(2014{\natexlab{b}})\citenamefont {Green}, \citenamefont
  {Engelhardt}, \citenamefont {Krieg}, \citenamefont {Negele}, \citenamefont
  {Pochinsky},\ and\ \citenamefont {Syritsyn}}]{Green:2012ud}%
  \BibitemOpen
  \bibfield  {author} {\bibinfo {author} {\bibfnamefont {J.~R.}\ \bibnamefont
  {Green}}, \bibinfo {author} {\bibfnamefont {M.}~\bibnamefont {Engelhardt}},
  \bibinfo {author} {\bibfnamefont {S.}~\bibnamefont {Krieg}}, \bibinfo
  {author} {\bibfnamefont {J.~W.}\ \bibnamefont {Negele}}, \bibinfo {author}
  {\bibfnamefont {A.~V.}\ \bibnamefont {Pochinsky}}, \ and\ \bibinfo {author}
  {\bibfnamefont {S.~N.}\ \bibnamefont {Syritsyn}},\ }\bibfield  {title}
  {\enquote {\bibinfo {title} {Nucleon structure from lattice {QCD} using a
  nearly physical pion mass},}\ }\href {\doibase
  10.1016/j.physletb.2014.05.075} {\bibfield  {journal} {\bibinfo  {journal}
  {Phys. Lett. B}\ }\textbf {\bibinfo {volume} {734}},\ \bibinfo {pages} {290}
  (\bibinfo {year} {2014}{\natexlab{b}})},\ \Eprint
  {http://arxiv.org/abs/1209.1687} {arXiv:1209.1687 [hep-lat]} \BibitemShut
  {NoStop}%
\bibitem [{\citenamefont {Morningstar}\ and\ \citenamefont
  {Peardon}(2004)}]{Morningstar:2003gk}%
  \BibitemOpen
  \bibfield  {author} {\bibinfo {author} {\bibfnamefont {C.}~\bibnamefont
  {Morningstar}}\ and\ \bibinfo {author} {\bibfnamefont {M.~J.}\ \bibnamefont
  {Peardon}},\ }\bibfield  {title} {\enquote {\bibinfo {title} {{Analytic
  smearing of $SU(3)$ link variables in lattice QCD}},}\ }\href {\doibase
  10.1103/PhysRevD.69.054501} {\bibfield  {journal} {\bibinfo  {journal} {Phys.
  Rev. D}\ }\textbf {\bibinfo {volume} {69}},\ \bibinfo {pages} {054501}
  (\bibinfo {year} {2004})},\ \Eprint {http://arxiv.org/abs/hep-lat/0311018}
  {arXiv:hep-lat/0311018} \BibitemShut {NoStop}%
\bibitem [{\citenamefont {Pochinsky}()}]{Qlua}%
  \BibitemOpen
  \bibfield  {author} {\bibinfo {author} {\bibfnamefont {A.}~\bibnamefont
  {Pochinsky}},\ }\href@noop {} {\enquote {\bibinfo {title} {Qlua},}\ }\bibinfo
  {howpublished} {\url{https://usqcd.lns.mit.edu/qlua}}\BibitemShut {NoStop}%
\bibitem [{\citenamefont {Babich}\ \emph {et~al.}(2010)\citenamefont {Babich},
  \citenamefont {Brannick}, \citenamefont {Brower}, \citenamefont {Clark},
  \citenamefont {Manteuffel}, \citenamefont {McCormick}, \citenamefont
  {Osborn},\ and\ \citenamefont {Rebbi}}]{Babich:2010qb}%
  \BibitemOpen
  \bibfield  {author} {\bibinfo {author} {\bibfnamefont {R.}~\bibnamefont
  {Babich}}, \bibinfo {author} {\bibfnamefont {J.}~\bibnamefont {Brannick}},
  \bibinfo {author} {\bibfnamefont {R.~C.}\ \bibnamefont {Brower}}, \bibinfo
  {author} {\bibfnamefont {M.~A.}\ \bibnamefont {Clark}}, \bibinfo {author}
  {\bibfnamefont {T.~A.}\ \bibnamefont {Manteuffel}}, \bibinfo {author}
  {\bibfnamefont {S.~F.}\ \bibnamefont {McCormick}}, \bibinfo {author}
  {\bibfnamefont {J.~C.}\ \bibnamefont {Osborn}}, \ and\ \bibinfo {author}
  {\bibfnamefont {C.}~\bibnamefont {Rebbi}},\ }\bibfield  {title} {\enquote
  {\bibinfo {title} {{Adaptive multigrid algorithm for the lattice Wilson-Dirac
  operator}},}\ }\href {\doibase 10.1103/PhysRevLett.105.201602} {\bibfield
  {journal} {\bibinfo  {journal} {Phys. Rev. Lett.}\ }\textbf {\bibinfo
  {volume} {105}},\ \bibinfo {pages} {201602} (\bibinfo {year} {2010})},\
  \Eprint {http://arxiv.org/abs/1005.3043} {arXiv:1005.3043 [hep-lat]}
  \BibitemShut {NoStop}%
\bibitem [{\citenamefont {Osborn}\ \emph {et~al.}()\citenamefont {Osborn} \emph
  {et~al.}}]{QOPQDP}%
  \BibitemOpen
  \bibfield  {author} {\bibinfo {author} {\bibfnamefont {J.}~\bibnamefont
  {Osborn}} \emph {et~al.},\ }\href@noop {} {\enquote {\bibinfo {title}
  {{QOPQDP}},}\ }\bibinfo {howpublished}
  {\url{https://usqcd-software.github.io/qopqdp/}}\BibitemShut {NoStop}%
\bibitem [{\citenamefont {{Jülich Supercomputing Centre}}(2019)}]{juwels}%
  \BibitemOpen
  \bibfield  {author} {\bibinfo {author} {\bibnamefont {{Jülich Supercomputing
  Centre}}},\ }\bibfield  {title} {\enquote {\bibinfo {title} {{JUWELS: Modular
  Tier-0/1 Supercomputer at the Jülich Supercomputing Centre}},}\ }\href
  {\doibase 10.17815/jlsrf-5-171} {\bibfield  {journal} {\bibinfo  {journal}
  {J. Large-Scale Res. Facil.}\ }\textbf {\bibinfo {volume} {5}},\ \bibinfo
  {pages} {A135} (\bibinfo {year} {2019})}\BibitemShut {NoStop}%
\end{thebibliography}%

\end{document}